\documentclass[twocolumn,secnumarabic,amssymb, nobibnotes, aps, prl]{revtex4-2}

\usepackage{bibunits}
\defaultbibliographystyle{apsrev4-2}  
\defaultbibliography{main}

\usepackage{braket}
\usepackage{physics}
\usepackage{mathtools}
\usepackage{amsfonts}
\usepackage{graphicx}
\usepackage{dcolumn}
\usepackage{bm}
\usepackage{diagbox}

\usepackage{float}

\usepackage{placeins} 

\newsavebox{\mstrut}
\newcommand{\sbra}[1]{%
    \sbox{\mstrut}{\(#1\)}%
    \mathinner{\langle\!\langle{#1}|}
}
\newcommand{\sket}[1]{%
    \sbox{\mstrut}{\(#1\)}%
     \mathinner{|{#1}\rangle\!\rangle}
}
\newcommand{\sbraket}[2]{%
    \sbox{\mstrut}{\(#1\)}%
    \sbox{\mstrut}{\(#2\)}%
   \mathinner{\langle\!\langle{#1}|{#2}\rangle\!\rangle}
}  
\newcommand{\sketbra}[2]{%
    \sbox{\mstrut}{\(#1\)}%
    \sbox{\mstrut}{\(#2\)}%
    \mathinner{|{#1}\rangle\!\rangle\langle\!\langle{#2}|}
}
\newcommand{\scom}[1]{%
    \sbox{\mstrut}{\(#1\)}%
    \mathinner{\left[\mkern-2mu\left[{#1}\right]\kern-0.1\ht\mstrut\right]}%
}
\newcommand{\vL}{\bar{\mathcal{L}}}

\begin{document}
\begin{bibunit}

\title{Liouville Fock state lattices and potential simulators}%

\author{Caio B. Naves}%
\email[]{caio.naves@fysik.su.se}
\affiliation{Department of Physics,
Stockholm University, AlbaNova University Center, 106 91 Stockholm,
Sweden}
\author{Jonas Larson}%
\email[]{jolarson@fysik.su.se}
\affiliation{Department of Physics, Stockholm University, AlbaNova University Center, 106 91 Stockholm, Sweden}
\begin{abstract}
We introduce Liouville Fock state lattices (LFSLs) as a framework for visualizing open quantum systems through matrix representations of the Lindblad master equation (LME). By vectorizing the LME, the state evolves in a doubled Hilbert space, naturally forming a synthetic lattice. Unlike the unitary evolution of pure states, LFSL states exhibit nontrivial dynamics due to the non-Hermitian Liouvillian, featuring population drifts, sources, and sinks—paralleling stochastic classical lattices. We explore these "classical simulators" in both the Fock representation and alternative positive semidefinite representations, which more closely resemble classical probability distributions. We further demonstrate how infinite steady state manifolds can derive from frustration in the LFSL.  
  
\end{abstract}

\maketitle

\textit{Introduction} -- Lattice models are essential tools for describing a wide range of physical phenomena across various branches of physics, including statistical mechanics, optics, condensed matter, and high-energy physics. A notable example is quantum simulators, which often rely on discretized lattice frameworks~\cite{bloch2008many, lewenstein2007ultracold}. The precise control achievable in ultracold atomic and ionic systems has made these simulators a practical reality. Over the past two decades, quantum simulators have enabled groundbreaking experiments, from realizing canonical condensed matter models~\cite{greiner2002quantum, britton2012engineered, hirthe2023magnetically} to exploring quantum thermalization and many-body localization~\cite{kaufman2016quantum, choi2016exploring, abanin2019colloquium}. Recently, quantum simulation has expanded further to incorporate photonic systems and address applications in high-energy physics and quantum chemistry~\cite{altman2021quantum}.

Building on the flexibility of these quantum systems, researchers have proposed methods for creating synthetic dimensions by exploiting internal degrees of freedom~\cite{boada2012quantum, celi2014synthetic, wang2016mesoscopic, ozawa2019topological, yuan2024quantum}. These synthetic dimensions allow for the study of phenomena unique to higher-dimensional systems, which are typically absent in traditional three-dimensional setups~\cite{yuan2018synthetic, yuan2021synthetic, ozawa2016synthetic, mccanna2021superfluid, bouhiron2024realization}. Synthetic dimensions have been proposed and experimentally realized in various platforms, including cold atoms~\cite{boada2012quantum, celi2014synthetic, price2017synthetic, mumford2022meissner, oliver2023bloch, reid2024topological, baum2018setting, yin2022floquet}, trapped ions~\cite{mancini2015observation, wang2024realizing}, magnons~\cite{price2020synthetic}, plasmonics~\cite{wu2024synthetic}, cavity and circuit QED systems~\cite{wang2016mesoscopic, deng2022observing, saugmann2023fock, cai2023nodal, larson2024floquet, yuan2024quantum,zhang2024synthetic}, electrical circuits~\cite{wang2020circuit}, and optical systems~\cite{perez2010glauber, lin2016photonic, tschernig2020multiphoton, ozawa2019topological, lustig2019photonic, dutt2020higher, dutt2020single, wu2023observation}.

Typically, these synthetic dimensions correspond to internal quantum states, forming what are known as Fock state lattices (FSLs). In many systems, particularly in quantum optics, such lattices lack translational invariance due to state-dependent tunneling rates. Nevertheless, essential properties such as topology are governed by the overall lattice geometry rather than by translational symmetry~\cite{deng2022observing, cai2021topological, wu2023observation, cai2023nodal, saugmann2023fock, yuan2024quantum}. This makes FSLs a powerful platform for exploring physical phenomena in state space rather than real space. While the FSL approach often reduces the problem to a single-particle framework, it has been experimentally demonstrated that computationally challenging problems can also be addressed using FSLs~\cite{sturges2021quantum}, further showcasing their potential for quantum simulation.

In this Letter, we extend the concept of FSLs to open quantum systems by introducing LFSLs derived from the LME. Extending phenomena from closed to open quantum systems presents unique challenges, particularly due to the role of quantum coherences in mixed states~\cite{streltsov2017colloquium}. These challenges are well recognized in areas such as entanglement and topological properties of mixed states~\cite{bennett1996mixed,sjoqvist2000geometric, heyl2017dynamical, lieu2020tenfold}, coherent control of open systems~\cite{koch2016controlling}, and defining thermodynamic quantities~\cite{kosloff2013quantum}. The difficulty arises from the fact that general density matrices $\hat{\rho}$ span a much larger state space than pure states $|\psi\rangle$. Unlike traditional FSLs, where vectors are represented in terms of probability amplitudes, the vectors in LFSLs require a fundamentally different physical interpretation. While these differences introduce additional complexities, they also offer new opportunities beyond Hermitian systems to explore novel phenomena, like anomalous transport~\cite{metzler2000random, zaslavsky2002chaos} and classical lattices/networks with biased complex weights~\cite{rothman1994lattice,harris2005current,fronczak2009biased,bottcher2024complex}. To demonstrate the power of this approach, we provide examples, including the emergence of frustration within the extended space of LFSLs.


\textit{Liouville-Fock-state lattices} -- Our approach is based on the LME~\cite{lindblad1976generators,breuer2002theory}, given by
\begin{equation}
    \frac{d\hat{\rho}}{dt} = \mathcal{L}\left[\hat{\rho}\right] \equiv -i\left[\hat{H}, \hat{\rho}\right] + \sum_j \gamma_j \left(\hat{L}_j \hat{\rho} \hat{L}_j^\dagger - \frac{1}{2} \left\{\hat{L}_j^\dagger \hat{L}_j, \hat{\rho}\right\}\right),
    \label{eq:lindblad}
\end{equation}
where $\mathcal{L}$ is the Liouvillian superoperator, $\hat{H}$ is the system Hamiltonian, $\hat{L}_j$ are the Lindblad jump operators, and $\gamma_j$ are the corresponding decay rates (we use hats on normal operators and calligraphic font for superoperators). The Liouvillian, being a superoperator that maps operators to other operators, makes it challenging to interpret the LME in terms of a FSL. To address this, we adopt the vectorized form of the LME.

This approach involves mapping the density matrix $\hat{\rho}$ to a "superstate" $\sket{\rho}$ by transforming bras into kets, such that $\ketbra{\psi_n}{\psi_m} \mapsto \ket{\psi_n} \otimes \ket{\psi_m}^*$. This mapping is a specific instance of the more general Choi–Jamiołkowski isomorphism~\cite{choi1975completely,jamiolkowski1972linear}. In this representation, the vectorized state $\sket{\rho}$ resides in the doubled Hilbert space $\mathfrak{L} = \mathfrak{H} \otimes \mathfrak{H}^*$, commonly referred to as the Liouville space~\cite{fano_liouville_1964,gyamfi_fundamentals_2020}. 

To proceed, we assume the existence of a natural number basis $\{|n\rangle;\, n \in \mathbb{N}\}$, corresponding to Fock states. For a single degree of freedom, the vectorized density matrix is then expanded as $\sket{\rho} = \sum_{n,m} \rho_{n,m} \sket{n,m}$, where the Fock basis for $\mathfrak{L}$ is $\{\sket{n,m}=|n\rangle|m\rangle^*\}$.

Within this vectorized framework, the LME is written as $\partial_t \sket{\rho} = \vL \sket{\rho}$, where the vectorized Liouvillian is given by
\begin{equation}
    \vL = -i\scom{\hat{H}, \hat{I}}_{-} + \sum_j \gamma_j \left(\hat{L}_j \otimes \hat{L}_j^* - \frac{1}{2} \scom{\hat{L}_j^\dagger \hat{L}_j, \hat{I}}_+\right),
    \label{eq:liouville}
\end{equation}
and the super-(anti)commutator is defined as $\scom{\hat{A}, \hat{B}}_{\pm} = \hat{A} \otimes \hat{B}^T \pm \hat{B} \otimes \hat{A}^T$. Here, $\hat{I}$ represents the identity operator. Throughout this work, we use an overhead bar to denote the vectorized form of superoperators, $\mathcal{S} \rightarrow \bar{\mathcal{S}}$. The matrix elements of the Liouvillian read
$\vL_{j,k}^{n,m}=\sbra{n,m}\vL\sket{j,k}$.

While the vector $\sket{\rho}$ belongs to the Hilbert space $\mathfrak{L}$, its physical interpretation differs fundamentally from that of pure states in the conventional Hilbert space $\mathfrak{H}$. For instance, the norm of $\sket{\rho}$, defined via the Hilbert-Schmidt inner product $\sbraket{A}{B} = \mathrm{Tr}(A^\dagger B)$, corresponds to the purity of the state, given by $\sbraket{\rho}{\rho} = \mathrm{Tr}(\rho^2) \leq 1$. Probability conservation for $\sket{\rho}$ is instead ensured by the condition $\sbraket{\mathbb{I}}{\rho} = 1$, where $\sbra{\mathbb{I}}$ represents the vectorized identity operator. Furthermore, physical expectation values in the vectorized formalism are computed as $\langle\hat{A}\rangle = \mathrm{Tr}(\hat{A}\hat{\rho}) = \sbraket{A}{\rho}$. Note also, that in this framework, the components of $\sket{\rho}$ correspond to the ``populations" of individual lattice sites in the LFSL. Since these components are generally complex, they cannot represent valid populations at each site, even though we will refer to them as such for convenience. 

It should be clear that the matrix $\vL$ defines the LFSL in the same way the Hamiltonian defines the regular FSLs. For a finite-dimensional Hilbert space $\mathfrak{H}$ with $\mathrm{Dim}(\mathfrak{H}) = D$, the corresponding Liouville space $\mathfrak{L}$ has dimension $\mathrm{Dim}(\mathfrak{L}) = D^2$. Since $\vL$ is not a Hermitian matrix, the evolution it generates is not unitary. Determining the lattice structure from $\vL$, such as its dimensions and bond identifications, is not unique. However, in many cases, there exists a natural choice, which typically restricts tunneling to nearby lattice sites.  

The LME defines a dynamical semigroup, where each element is a completely positive and trace-preserving (CPTP) map. This ensures that any physical state $\hat{\rho}$ (i.e., a positive semi-definite operator with unit trace) remains a physical state under the LME's evolution. 
While the CPTP property is not a symmetry in the conventional sense~\cite{kim_third_2023}, it imposes dynamical constraints on the system, including a detailed balance condition expressed as
\begin{equation}
    \sum_{n} \bigg(\frac{d}{dt}\rho_{n,n} - \sum_{j,k} \vL_{j,k}^{n,n} \rho_{j,k}\bigg) = 0,
\end{equation}
which ensures that the sum of the changes in population is always balanced by the Liouvillian dynamics.


\textit{Symmetries and conserved quantities} -- For Hamiltonian systems, Noether's theorem establishes a fundamental connection between conserved quantities and continuous symmetries. In the context of FSLs, symmetries play a crucial role in determining the structure of the lattice, as explored in detail in~\cite{saugmann2023fock}. However, in open quantum systems, the direct correspondence implied by Noether's theorem does not generally hold. Nonetheless, symmetries still have a significant impact on the geometry of the LFSL.

According to the definitions in Ref.~\cite{albert_symmetries_2014}, symmetries in open systems can be classified into two categories: weak and strong symmetries. A weak symmetry refers to a symmetry of the Liouvillian superoperator that does not necessarily correspond to a conserved quantity. To formalize this, let $\hat{S}$ denote the Hermitian generator of a continuous symmetry of the Liouvillian. Specifically, the symmetry condition is expressed as $\mathcal{U}^\dagger \mathcal{L} \mathcal{U} = \mathcal{L}$, where $\mathcal{U} = \exp(i\phi \mathcal{S})$ and $\mathcal{S}$ is the superoperator representation of $\hat{S}$. It can be further shown that the following holds $[\hat{S}, \hat{H}] =0$ and $[\hat{S},\hat{L}_j]=\alpha_j\hat L_j$ with $\alpha_j \in \mathbb{R}$~\cite{albert_symmetries_2014}. If $\alpha_j=0$ for all $j$, i.e. $[\hat{S}, \hat{H}] = [\hat{S}, \hat{L}_j] = 0$, the symmetry is classified as strong. In this case, the symmetry not only applies to the Liouvillian but also directly corresponds to a conserved quantity. 

The discussion of conserved quantities is most naturally framed in the Heisenberg picture, where the time evolution of an operator $\hat{S}$ is given by $d\hat{S}/dt = \mathcal{L}^\dagger(\hat{S})$. In the vectorized formalism, a conserved quantity in open dynamics corresponds to an operator $\hat{S}$ that satisfies the condition $\vL^{\dagger}\sket{S} = 0$. We now make the following observation. Since $\vL$ and its adjoint $\vL^{\dagger}$ share the same spectrum, if $\vL$ has $d$ linearly independent steady states, i.e., $\vL\sket{\psi_i} = 0 \mbox{ for } i = 1,\dots,d$, then $\vL^\dagger$ must also have $d$ linearly independent steady states, satisfying $\vL^\dagger\sket{S_i} = 0 \mbox{ for } i = 1,\dots,d$. It then follows that these states $\sket{S_i}$ correspond to conserved observables. This can be seen by evaluating  
\[
\sbraket{S_i}{\rho_{ss}}\! =\! \lim_{t \rightarrow \infty}\!\sbraket{\!S_i}{e^{\vL t}\rho_{in}\!}\! =\! \lim_{t \rightarrow \infty}\!\sbraket{\!e^{\vL^\dagger t}S_i}{\rho_{in}\!}\! =\! \sbraket{S_i}{\rho_{in}}\!,
\] 
where $\rho_{ss}$ is a steady state and $\rho_{in}$ is the initial state. This result, known as the steady-state conserved quantity correspondence~\cite{albert_symmetries_2014}, demonstrates that conserved observables $\hat{S}_i$ encode information about the initial conditions of the system.

Returning to the LFSL formalism, a continuous symmetry (whether weak or strong), as in the case of FSLs, leads to a splitting of the lattice into disconnected sublattices, each with a dimension reduced by one. In contrast, a discrete symmetry causes the lattice to break into distinct sublattices but without any reduction in dimensionality. However, a key distinction between FSLs and LFSLs is that in LFSLs it is possible to have sublattices where the corresponding populations vanish. These sublattices represent decaying subspaces and arise as a result of weak symmetries. For the populations to remain preserved, the lattice splitting must instead originate from strong symmetries that are associated with conserved quantities.


\textit{Examples} -- As a first example, we consider a single cavity mode with Hamiltonian $\hat{H} = \omega \hat{a}^\dagger \hat{a}$ undergoing two-photon loss, described by the jump operator $\hat{L} = \hat{a}^2$. Since the Liouville space is given by the direct product of two bosonic Hilbert spaces, making the LFSL two-dimensional, where sites are labeled by the Fock numbers $n$ and $m$; $\sket{n,m}$. 

This system possesses a weak symmetry generated by the number operator $\hat{N} = \hat{a}^\dagger \hat{a}$~\cite{albert_symmetries_2014}, which leads to a splitting of the lattice into one-dimensional sublattices characterized by a fixed integer difference $\Delta = n - m$. Additionally, the system exhibits a discrete strong symmetry corresponding to the parity transformation $\hat{a} \rightarrow -\hat{a}$. This symmetry further divides the one-dimensional sublattices into two sets, each associated with a parity sector of even or odd of $\mathfrak{H}$ and $\mathfrak{H}^*$. 

\begin{figure}[!ht]
    \centering
    \includegraphics[width=0.6\linewidth]{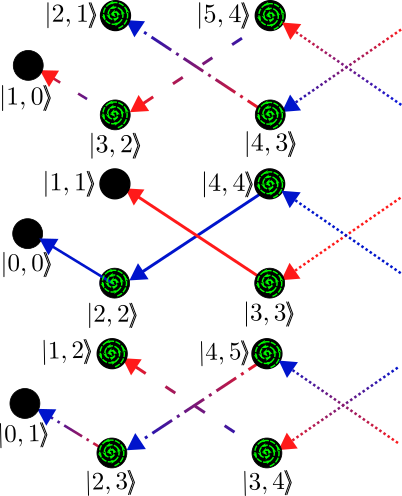}
    \caption{The LFSL corresponding to a harmonic oscillator undergoing two-photon loss. The depicted lattice is restricted to the diagonal and the first two off-diagonal elements of the density matrix. Full blue and red directed arrows indicate transitions between states with even and odd parities in $\mathfrak{H}\otimes\mathfrak{H}^*$, respectively, the dot-dashed lines represent transitions between even states on $\mathfrak{H}$ and odds in $\mathfrak{H}^*$ while the long dashed ones the transitions between odd states on $\mathfrak{H}$ and even on $\mathfrak{H}^*$. The onsite green spiral arrows represent decaying states.}
    \label{fig:two_photon_loss}
\end{figure}

Figure~\ref{fig:two_photon_loss} illustrates this structure, where blue and red arrows indicate parity and hopping direction, while sites marked with green spirals represent onsite decays. The four decay-free sites span the steady-state subspace. Among these, $\sket{0,0}$ and $\sket{1,1}$ correspond to the zero- and one-photon Fock states, while the remaining two represent their coherences. 

Introducing a coherent drive $\eta(\hat{a}^\dagger + \hat{a})$ in the Hamiltonian or incorporating single-photon loss would couple the parity sectors, significantly altering the structure and dynamics of the LFSL.

As a second example, we consider a model that results in a translationally invariant LFSL, namely a single particle hopping, coherently or incoherently, between neighboring sites in a one-dimensional lattice. What makes this model interesting is that while the Hamiltonian is trivial, in the Liouville space the model is frustrated. The Hamiltonian is given by  
\begin{equation}\label{tbh}
\hat{H} = \eta\sum_{n}(\ketbra{n}{n+1} + \text{h.c.}),
\end{equation}
and we introduce the jump operators  
\begin{equation}\label{hop}
    L_1 = \sum_{n}\ketbra{n}{n+1}, \quad L_2 = \sum_{n}\ketbra{n+1}{n},
\end{equation}
with corresponding rates $\gamma_1$ and $\gamma_2$, and $\ket{n}$ the state for a particle at site $n$. Since the jump operators commute with the Hamiltonian, $[\hat{L}_1,\hat{H}]=[\hat{L}_2,\hat{H}]=0$, we expect an energy dephasing, but the fact that $\hat{L}_{1,2}$ are non-hermitian makes the model non-trivial. The resulting LFSL forms a square lattice structure, as illustrated in Fig.~\ref{fig:lattice_model}, with red and blue lines representing incoherent and coherent tunnelings.

\begin{figure}[!ht]
    \centering
    \includegraphics[width=0.7\linewidth]{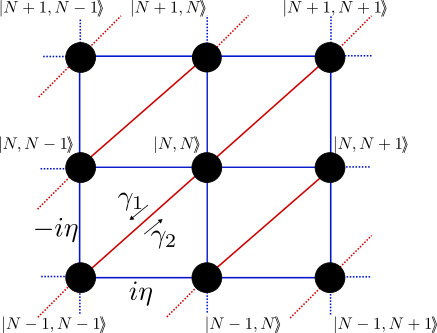}
    \caption{LFSL of the tight-binding model~(\ref{tbh}) with incoherent tunneling processes~(\ref{hop}). Horizontal and vertical blue lines arise from coherent tunneling, while diagonal red hoppings stem from the incoherent jumps.}
    \label{fig:lattice_model}
\end{figure}

Using the translational invariance of the LFSL, we can construct the eigenvectors of the Liouvillian using Bloch states $\ket{\theta}$, characterized by the quasi-momentum $\theta \in [-\pi/2, \pi/2)$. Specifically, the eigenstates take the form $\hat{\rho}_{\theta\tilde{\theta}} = |\theta\rangle\langle\tilde{\theta}|$. The steady states correspond to the diagonal elements $\hat{\rho}_{\theta\theta}$, and the spectrum of the Liouvillian is given by~\cite{SuppMat}
\begin{equation}
\begin{array}{lll}
    \mu(\theta,\tilde{\theta}) & = & -\left(\gamma_1+\gamma_2\right)\left[1-\cos\left(\theta-\tilde{\theta}\right)\right]\\ \\ 
    & & + i\left[\left(\gamma_1-\gamma_2\right)\sin\!\left(\!\theta-\tilde{\theta}\!\right)\! -\! 2\eta\!\left(\cos\theta-\cos\tilde{\theta}\right)\!\right]\!.
    \end{array}
\end{equation}

It is evident that the steady-state manifold, defined by $\theta = \tilde{\theta}$, forms a connected structure. As discussed further in~\cite{SuppMat}, this infinite degeneracy arises from geometric frustration. Similar to classical frustration~\cite{moessner2006geometrical}, this can be understood by considering a single unit cell and constructing a (energy) functional by assigning a complex order parameter to each site. In a frustrated system, it is not possible to simultaneously optimize all individual terms of the functional. Experimentally, such frustration manifests as distinctly different steady states across various realizations. Notably, this form of frustration emerges from the interplay between coherent and incoherent processes within the LME, and hence is unique to an open system.

\textit{Alternative representations} --
In the design of synthetic lattices, the choice of representation for vectorizing the system plays a crucial role, as it directly influences both the lattice structure and the dynamics involved. While Fock states are a natural choice due to their experimental relevance, they present challenges for simulating classical lattice dynamics, primarily because the resulting populations are generally complex. Similar to phase-space methods~\cite{gardiner2004quantum}, alternative lattice representations of $\hat{\rho}$ exist that ensure real-valued and non-negative populations.

One such alternative is the Bloch representation, where the state is expressed as~\cite{kimura2003bloch,bertlmann2008bloch}
\begin{equation}
     \hat{\rho} = \frac{1}{d} \left( \hat{I} + \sqrt{\frac{d(d-1)}{2}} \mathbf{R} \cdot \vec{\lambda} \right),
\end{equation}
where $d$ is the Hilbert space dimension, $\mathbf{R} = (R_1, R_2, \dots, R_{d^2-1})$ is the Bloch vector, and $\vec{\lambda} = (\hat{\lambda}_1, \hat{\lambda}_2, \dots, \hat{\lambda}_{d^2-1})$ is a vector composed of the generalized Gell-Mann matrices~\cite{hioe1981n}. 

Although the Bloch vector satisfies $|\mathbf{R}| \leq 1$ and remains real, its components can take negative values. Moreover, for $d > 2$, only a subset of the Bloch sphere corresponds to physically valid quantum states. By extending the Bloch vector to include $R_0 = 1$ and defining an augmented basis with $\hat{\lambda}_0 = \hat{I}$, the LME takes the compact homogeneous form 
\begin{equation}
    \frac{d\mathbf{R}}{dt} = \hat{\mathfrak{L}} \mathbf{R},
\end{equation}
where the generator $\hat{\mathfrak{L}}$ is unitarily related to $\bar{\mathcal{L}}$, i.e.  $\hat{\mathfrak{L}} = \hat{V} \bar{\mathcal{L}} \hat{V}^\dagger$ for a unitary $\hat{V}$~\cite{SuppMat}.

A representation that ensures positive semi-definiteness is based on symmetric informationally complete positive operator-valued measures (SIC-POVMs)~\cite{renes_sicpovm_2004}. To construct this representation, we define a complete set of projectors $\hat{\Pi}_i = (1/d)\ketbra{\psi_i}$, which satisfy the constant overlap condition $\mathrm{Tr}\left[\hat{\Pi}_i\hat{\Pi}_j\right] = \frac{1}{d^2} \frac{d\delta_{i,j} + 1}{d + 1}$. Using this basis, the density matrix can be expanded as  
\begin{equation}
    \sket{\rho} = \sum_{i = 1}^{d^2} \bigg( d(d+1)p_i - 1 \bigg)\sket{\Pi_i},
\end{equation}  
where $\sket{\Pi_i}$ are the vectorized elements of the POVM, and $p_i = \mathrm{Tr}\left[\hat{\rho}\hat{\Pi}_i\right] \geq 0$ represents the probability of finding the system in the state $\sket{\Pi_i}$. Within this representation, the probability vector $\vec{p}$ evolves as  
\begin{equation}
    \frac{d \vec{p}}{dt} = M\vec{p} - \vec{m},
\end{equation}  
where the matrix elements $M_{i,j}$ are given by $M_{i,j} = d(d+1)\sbra{\Pi_i}\vL\sket{\Pi_j}$, and the components of $\vec{m}$ are defined as $m_i = \sum_{j} M_{i,j}/d(d+1)$.  

Interpreting this as a lattice model, the inhomogeneous term $\vec{m}$ acts as a source/sink term, reflecting the openness of the system. This becomes evident from the evolution equation for the components, $ \dot{p}_i = \sum_j M_{i,j} (p_j - 1)$,
where the inhomogeneous term behaves as a constant, non-local background field that can create or annihilate population. In analogy with classical stochastic master equations, probability conservation in the homogeneous case is ensured by the constraint $\sum_{i} M_{i,j} = 0$~\cite{van1992stochastic}. In such cases, the negative terms of $M$ typically reside on its diagonal, representing transitions out of these sites. However, in our models negative terms can also appear on off-diagonal elements, arising from prohibited transitions—potentially due to quantum interference—that induce a preferred drift in the lattice. For an example of the lattice structure generated by $M$, see~\cite{SuppMat}.  


\textit{Classical simulations} -- The lattice models arising from vectorizing the LME typically result in drifts, sources, and sinks, which are more commonly encountered in classical models. This suggests potential applications for using open quantum systems to simulate classical lattice/network models. While the Fock and Bloch representations do not directly reproduce classical probability distributions, they are nonetheless promising candidates for exploring generic  ~\cite{bottcher2024complex}. A characteristic feature of these models is, for example, anomalous diffusion/transport~\cite{metzler2000random, zaslavsky2002chaos}, an active area of research due to its applications in various fields such as biology~\cite{hofling2013anomalous} and pollution spread~\cite{ganti2010normal}.

The SIC-POVM representation of the density matrix is positive semi-definite; however, as mentioned earlier, its lattice evolution may involve negative tunneling rates. These negative rates typically result in anomalous transport in the lattice, linking the system yet again to the aforementioned classical models. 

\textit{Conclusion} -- In summary, we present LFSLs as an extension of FSLs to open quantum systems. By vectorizing the LME, we map the system onto an extended Hilbert space, naturally forming a synthetic lattice with distinct structural and dynamical properties. Unlike conventional FSLs, the LFSLs are governed by non-Hermitian dynamics, leading to population drifts, sources, and sinks, and requiring a reinterpretation of state vectors beyond probability amplitudes. Furthermore, the role of symmetries in LFSLs deviates from Noether’s theorem in closed systems, fundamentally altering their geometric and topological characteristics. Given these unique features, we suggest that LFSLs provide a promising framework for simulating non-equilibrium phenomena such as anomalous transport in stochastic classical lattice models. A detailed analysis of such applications is left for the future.

\begin{acknowledgements}
The authors thank Thomas Klein Kvorning, Supriya Krishnamurthy, and Yu Tian for fruitful and insightful discussions. 
\end{acknowledgements}

\end{bibunit}

\clearpage 
\onecolumngrid

\begin{bibunit}

\section{Supplementary material: Liouville Fock state lattices and simulators}%

\section*{Representations of $\hat{\rho}$}
In this section, we provide a more detailed discussion of various representations of the density operator. In the main text, we introduce three such representations, with particular emphasis on the Fock representation, also known as the Liouville representation. This choice is natural, as it offers several advantages. First, once the model is specified, identifying the LFSL is relatively straightforward. Second, the Fock representation is often experimentally relevant, as Fock states commonly correspond to eigenstates of the bare Hamiltonian. Third, since the LME is typically formulated in terms of creation and annihilation operators, this basis facilitates the identification of symmetries and conserved quantities. However, when simulating classical lattice or network models, this representation presents a challenge due to its use of complex-valued `probabilities.' This motivates the consideration of alternative, purely real representations, even though key aspects of, say, anomalous transport can likely still be studied within the Liouville framework.  

One such alternative is the Bloch representation, which can be derived from the Liouville representation via a unitary transformation. This transformation effectively separates the real and imaginary components of the population elements, resulting in a fully real representation, although some elements may still take negative values. Another option is the SIC-POVM representation, which is not only purely real but also strictly non-negative, making it resemble a proper probability distribution. However, this approach introduces negative tunneling rates, and extending it to higher-dimensional lattices becomes increasingly complicated.

\subsection*{The Liouville representation - vectorization and Liouville-Fock State Lattices}
The Fock or Liouville representation of the density operator $\hat{\rho}$ is constructed using the bra-flipper isomorphism, denoted by the operator \rotatebox[origin=c]{180}{$\Omega$}. This isomorphism maps any density matrix $\hat{\rho}$ to a corresponding superstate, or \textit{superket}, $\sket{\rho}$, which resides in the Liouville space $\mathfrak{L} = \mathfrak{H} \otimes \mathfrak{H}^*$~\cite{fano_liouville_1964, gyamfi_fundamentals_2020}. Consider a Hilbert space $\mathfrak{H}$ with an orthonormal basis $\{\ket{\psi_n}, n \in \mathbb{N}\}$. The space of bounded operators $\mathcal{B}(\mathfrak{H})$ can then be expanded using the basis elements $\{\ketbra{\psi_n}{\psi_m}, n,m \in \mathbb{N}\}$. Under vectorization, these elements transform according to the rule  
\begin{equation}
    \ketbra{\psi_n}{\psi_m} \rightarrow \ket{\psi_n} \otimes \ket{\psi_m}^{*} = \sket{\psi_n, \psi_m},
\end{equation}  
which effectively maps $\mathfrak{H}$ to $\mathfrak{H} \otimes \mathfrak{H}^*$. Importantly, the state $\sket{\psi_n, \psi_m}$ is generally distinct from the standard tensor product state $\ket{\psi_n, \psi_m} \in \mathfrak{H} \otimes \mathfrak{H}$.  

A vectorized matrix $\hat{M} \in \mathcal{B}(\mathfrak{H})$ is referred to as a \textit{superket} and is denoted by $\sket{M} \in \mathfrak{L}$. Similarly, superoperators $S \in \mathcal{S}(\mathfrak{H})$, which act on operators in $\mathcal{B}(\mathfrak{H})$, are represented as matrices in this framework. The precise form of a vectorized superoperator depends on its action, but a key identity that facilitates calculations is the \textit{superket triple product identity}~\cite{gyamfi_fundamentals_2020}, given by  
\begin{equation}
    \rotatebox[origin=c]{180}{$\Omega$}(ABC) = (A \otimes C^T)\sket{B} = (AB \otimes I)\sket{C} = (I \otimes (BC)^T)\sket{A},
\end{equation}  
where $A^T$ denotes the matrix transpose of $A$. Applying this identity, the Liouvillian $\mathcal{L}$ can be rewritten as the non-Hermitian matrix  
\begin{equation}
    \vL = -i\scom{\hat{H}, \hat{I}}_{-} + \sum_j \gamma_j\big(\hat{L}_j\otimes \hat{L}_j^* - \frac{1}{2}\scom{\hat{L}_j^{\dagger}\hat{L}_j, \hat{I}}_+\big),
    \label{eq:liouville}
\end{equation}  
where the \textit{super (anti)commutator} is defined as  
\begin{equation}
    \scom{\hat{A},\hat{B}}_{\pm} = \hat{A} \otimes \hat{B}^T \pm  \hat{B} \otimes \hat{A}^T \,.
\end{equation}  
In this way, the vectorized density matrix evolves according to the following vector equation 
    \begin{equation}
        \frac{d}{dt}\sket{\rho} = \vL \sket{\rho}\mbox{.}
        \label{eq:vec_evolution}
    \end{equation}
As mentioned in the main text, we denote the vectorized form of a superoperator using an overbar, i.e., $\mathcal{S} \rightarrow \bar{S}$. 

Although the state vector $\sket{\rho}$ is formally a vector in a Hilbert space, its interpretation differs fundamentally from that of a standard quantum state vector $\ket{\psi} \in \mathfrak{H}$. Its norm, defined via the Hilbert-Schmidt inner product as  
\begin{equation}
    P \equiv \sbraket{\rho}{\rho} = \mbox{Tr}(\hat{\rho}^2) \leq 1,
\end{equation}  
corresponds to the \textit{purity} of the state, which quantifies the degree of mixedness. Unlike the norm of a wavefunction $\langle\psi|\psi\rangle$, which remains invariant under unitary evolution, the purity $P$ evolves dynamically under a given LME. However, the preservation of the trace of $\hat{\rho}$ can be expressed in the vectorized formalism as  
\begin{equation}
    \frac{d}{dt}\sbraket{I}{\rho} = \frac{d}{dt} \mbox{Tr}(\hat{\rho}) = 0,
    \label{eq:trace_preservation}
\end{equation}  
where $\sket{\mathbb{I}}$ denotes the vectorized identity operator. This equation implies that $\sket{\mathbb{I}}$ is a left-eigenvector of the Liouvillian with eigenvalue zero, i.e.,  
\begin{equation}
    \sbra{\mathbb{I}}\vL = 0.
\end{equation}  
The components of a pure state $\ket{\psi}$ in a given basis represent probability amplitudes, whereas the elements of $\sket{\rho}$ serve a different purpose. Since $\sket{\rho}$ is a vectorized form of $\hat{\rho}$, they contain the same elements, meaning that the components of $\sket{\rho}$ encode both populations (the diagonal elements of $\hat{\rho}$) and coherences (the off-diagonal elements of $\hat{\rho}$).  

So far, we have not specified a basis in which to analyze the vectorized Liouvillian $\vL$. Let us now introduce a natural number basis $\{\ket{n}\}$ for the Hilbert space $\mathfrak{H}$, which may for example correspond to bosonic or fermionic Fock states. In the bosonic case, these states satisfy $\hat{a} \ket{n} = \sqrt{n} \ket{n-1}$, while for fermions, the possible states are $\ket{0}$ and $\ket{1}$. For spin systems we can use the eigenstates of $\hat{S}_z$, given by $\hat{S}_z \ket{S,m} = m\ket{S,m}$ as our number states.  

In the Liouville space $\mathfrak{L}$, the corresponding Fock states are given by $\{\sket{n,m} \mid n,m \in \mathbb{N}\}$. Here, it should be understood that $n$ and $m$ can represent multiple modes and may include combinations of different particle types (bosonic, fermionic, or spin). Once we establish the Fock state basis $\{\sket{n,m}\}$, the matrix representation of $\vL$ in this basis defines the \textit{Liouville-Fock space lattice} (LFSL).  

For a finite-dimensional Hilbert space $\mathfrak{H}$ of dimension $d$, the Liouville space has dimension $D(\mathfrak{L}) = d^2$. Furthermore, the number of degrees of freedom is effectively doubled in Liouville space, meaning that the LFSL typically has twice the dimension of the corresponding Hamiltonian FSL. This feature will be illustrated through several examples in the following sections.  

The completely positive trace-preserving (CPTP) property of the LME requires that the Liouvillian $\vL$ preserves hermiticity of $\hat{\rho}$. As a consequence, the distribution on the lattice exhibits symmetry, up to complex conjugation, around the "diagonal" sites. Expanding the density matrix in the Fock basis as  
\begin{equation}
    \sket{\rho} = \sum_{n,m} \rho_{n,m} \sket{n,m},
\end{equation}  
this condition enforces the relation $\rho_{n,m} = \rho_{m,n}^*$. We refer to the coefficients $\rho_{n,m}$ as populations, even though they can, in general, be complex.  

The hermiticity-preserving property of $\vL$ can also be interpreted as a form of symmetry—though not a conventional real-valued one—as discussed in \cite{kim_third_2023}. Additionally, trace preservation manifests dynamically in the LFSL as a form of detailed balance for the sites corresponding to populations. Expressing the Liouvillian in the Fock basis, we write  
\begin{equation}
    \vL = \sum_{n,m,j,k} \vL_{j,k}^{n,m} \sketbra{n,m}{j,k}.
\end{equation}  
Using Eq.~(\ref{eq:trace_preservation}), we obtain  
\begin{equation}
    \sum_{n} \bigg(\frac{d}{dt} \rho_{n,n} - \sum_{j,k} \vL_{j,k}^{n,n} \rho_{j,k} \bigg) = 0,
\end{equation}  
which expresses that the total change in populations is always balanced by the action of the Liouvillian.  

\newpage


\subsection*{The Bloch representation}
\label{supm:bloch_rep}
The Liouville representation offers several advantages, such as its experimental relevance and the ease with which LFSLs can be extracted. However, it also presents disadvantages, particularly in the simulation of other lattice models. Since the elements $\rho_{n,m}$ of $\sket{\rho}$ represent populations in the lattice, it would be preferable if they were both real and non-negative.  

A straightforward way to transition from the Liouville-Fock representation to a real-valued representation is by utilizing the hermiticity of $\hat{\rho}$. Specifically, the combinations 
\begin{equation}
    \rho_{n,m} + \rho_{m,n} \quad \text{and} \quad i (\rho_{n,m} - \rho_{m,n})  
\end{equation}  
are both real quantities. This transformation defines the \textit{Bloch representation}, which ensures that the density matrix elements are expressed in terms of real numbers.  

It is known that the density matrix of any finite-dimensional system can be expanded in terms of the generalized Gell-Mann matrices $\hat{\lambda}$~\cite{bertlmann2008bloch}. This expansion takes the form  
\begin{equation}
    \rho = \frac{1}{d} \left(\mathbb{I} + \sqrt{\frac{d(d-1)}{2}} \vec{R} \cdot \vec{\lambda} \right),
\end{equation}  
where $\vec{R}=(R_1,\dots,R_{d^2-1})$ is the Bloch vector. Applying the bra-flipper operator, one obtains the vectorized form  
\begin{equation} 
    \sket{\rho} = \sum_{j = 0}^{d^2 - 1} \tilde{R}_j \sket{\Lambda_j},
\end{equation}  
with the coefficients  
\begin{equation}
    \tilde{R}_0 = \frac{1}{\sqrt{d}}, \hspace{1cm} \tilde{R}_j = \frac{\beta}{d} R_j \quad \text{for } j \neq 0.
\end{equation}  
Here, we define the superkets as $\sket{\Lambda_0} = \sket{\lambda_0} / \sqrt{d}$ and $\sket{\Lambda_i} = \sket{\lambda_i} / \sqrt{2}$ for $i \neq 0$, where $\lambda_0 = \mathbb{I}$ is the identity matrix.  
The (generalized) Bloch vector $\vec{\Tilde{R}} = (\Tilde{R}_0, \dots, \Tilde{R}_{d^2-1})^T$ evolves in time according to  
\begin{equation}
    \frac{d \vec{\Tilde{R}}}{dt} = \vL^{\Lambda} \vec{\Tilde{R}},
\end{equation}  
where the transformed Liouvillian is given by  
\begin{equation}
    \vL^{\Lambda} = V \vL V^\dagger, \quad \text{with} \quad V = \sum_{i = 0}^{d^2 - 1} \sket{i} \sbra{\Lambda_i}.
\end{equation}  
That is, $\vL^{\Lambda}$ is obtained via a basis transformation of the Liouvillian matrix. Consequently, given the Liouvillian $\vL$ in the Fock basis, it is, in principle, straightforward to compute $\vL^\Lambda$, from which the Bloch-representation lattice can be extracted.


\subsection*{The Urgleichung -- SIC-POVM representation}
\label{supm:sic_rep}
Another representation is provided by Symmetric Informationally Complete POVMs (SIC-POVMs)~\cite{renes_sicpovm_2004}. This approach has the distinct advantage of being both real and non-negative, thereby possessing the desired properties of a probability distribution. However, as noted in the main text, the matrix governing time evolution exhibits properties that are not typically found in classical stochastic master equations—most notably, the presence of negative transition rates. Moreover, in higher dimensions $d$, extracting the distribution and corresponding lattice become extremely difficult.

Defining a SIC-POVM consists in assigning a complete set of projectors $\Pi_i = (1/d)\ketbra{\psi_i}$, i.e. $\sum_i \Pi = \mathbb{I}$ with a constant overlap Tr$(\Pi_i\Pi_j) = (1/d^2)((d\delta_{i,j} + 1)/(d + 1)$. This property makes them interesting for quantum state tomography~\cite{busch_informationally_1991, chiribella_covariant_2004} as it optimizes the number of measurements needed for the task. One can show that the density matrix can be expanded as 
\begin{equation}
    \sket{\rho} = \sum_{i = 1}^{d^2} \bigg(d(d+1)p_i -1\bigg)\sket{\Pi_i}\mbox{,}
\end{equation}
with $\sket{\Pi_i}$ being the vectorized elements of the POVM and $p_i = \mbox{Tr}(\rho\Pi_i) \ge 0$ the probability (i.e. $\sum_ip_i=1$) to find the system in the state $\sket{\Pi_i}$. By using this expansion and projecting the evolution equation~(\ref{eq:vec_evolution}) onto the SIC-POVM basis we derive an inhomogeneous differential evolution equation for the probability vector $\vec{p}=(p_1,\dots,p_{d^2})$
\begin{equation}
    \frac{d \vec{p}}{dt} = d(d+1)M\vec{p} - \vec{m}\mbox{,}
    \label{eq:sicpovm_prob_vec_ev}
\end{equation}
where $M$ is a matrix whose elements are $M_{i,j} = \sbra{\Pi_i}\vL\sket{\Pi_j}$ and the components of the vector $\vec{m}$, $m_i = \sum_{j}M_{i,j}$. The similarity to a classical stochastic evolution equation with transition matrix $M$ and a source/sink term $\vec{m}$ is noticeable, however there are crucial differences that we discuss in the following.

In a classical stochastic equation, probability conservation is ensured by the constraint on the time-evolution generator, $\sum_{i}M_{i,j} = 0$, where $M_{i,j}$ are the matrix elements of $M$. That this property holds also here becomes evident by rewriting Eq.~(\ref{eq:sicpovm_prob_vec_ev}) in terms of the components of $\vec{p}$  
\begin{equation}
    \frac{d p_i}{dt} = \sum_{j = 0}^{d^2}\left[d(d+1)p_j - 1\right]M_{i,j}\mbox{,}
    \label{eq:sicpovm_prob_ev}
\end{equation}
Summing over $i$ confirms that $\sum_{i}M_{i,j} = 0$. Using that the completeness of the projectors in the vectorized formalism and noting that the identity operator is always conserved under Liouvillian evolution, $\vL^\dagger\sket{\mathbb{I}} = 0$, it follows that  
\begin{equation}
    \sum_i M_{i,j} = \sum_i \sbra{\Pi_i}\vL\sket{\Pi_j} = \sbra{\mathbb{I}}\vL\sket{\Pi_j} = 0\mbox{.}
    \label{eq:consv_prob}
\end{equation}
The crucial difference between $M$ and a stochastic matrix is the fact that it can have negative off-diagonal entries. This is an essential feature for an evolution that displays a periodic time-asymptotic behavior rather than a steady state one, as we will exemplify later.

The inhomogeneous term, represented by the vector $\vec{m}$, given that the elements of $M$ can be negative, act as a source or sink term for the components of the probability vector. Thus, it plays a crucial role in shaping the time-asymptotic behavior of the system. If the matrix $M$ is invertible, a unique steady-state solution exists and is given by  
\begin{equation}
    \vec{p}_{ss} = \frac{M^{-1}\vec{m}}{d(d+1)}\mbox{.}
    \label{eq:minv_ss}
\end{equation}
However, invertibility is not guaranteed, since $M$ is not an Hermitian matrix in general
\begin{equation}
    M_{i,j}^* = (\sbra{\Pi_i}\vL\sket{\Pi_j})^{\dagger} = \sbra{\Pi_j}\vL^\dagger\sket{\Pi_i} \ne M_{j,i}\mbox{.}
\end{equation}
Then one has that the steady state probability distribution has to satisfy
\begin{equation}
    p_i^{\mbox{ss}}= \frac{m_i - \sum_{j \ne i}d(d+1)M_{i,j}p_j^{\mbox{ss}}}{d(d+1)M_{i,i}}\mbox{.} 
    \label{eq:sicpovm_ss_cond}
\end{equation}

We can envision the case where $\vec{m} = 0$. For that to be true we have that  
\begin{equation}
    m_i = \sum_{j} M_{i,j} = \sbra{\Pi_i}\vL\sket{\Pi_j} = \sbra{\Pi_i}\vL\sket{\mathbb{I}} = 0\mbox{,}
    \label{eq:maxi_mixed_ss}
\end{equation}
where we have used the completeness property of the SIC-POVMs. This equation implies that the source term is zero when the maximally mixed state is a steady state of the dynamics, since $\vL\sket{\mathbb{I}} = 0$. Interestingly, this aligns with the condition for a classical stochastic process to converge to a uniform probability distribution as its steady state~\cite{van1992stochastic}, despite $M$ not necessarily being symmetric.  From the definition of probabilities in the SIC-POVM representation, $p_i = \mbox{Tr}(\Pi_i \rho)$, and using the fact that $\mbox{Tr}(\Pi_i) = 1/d$, it follows that for the maximally mixed state $\rho = (1/d)\mathbb{I}$, we have 
\begin{equation}
p_i = \frac{1}{d^2}, \quad \forall i.
\end{equation}  
This demonstrates that the probability distribution associated with the maximally mixed state is uniformly distributed, similar to the classical case. 

The above observation is merely a consequence of the correspondence between steady states in the present LME framework and those in SIC-POVM representation. Let $\{\sket{\rho_i}\}$ be the set of right eigenvectors of the Liouvillian, i.e.,  
\begin{equation}
    \vL \sket{\rho_i} = \lambda_i\sket{\rho_i}, \quad \lambda_i \in \mathbb{C}.
\end{equation}
Furthermore, let each eigenvector be expanded in the SIC-POVM basis as  
\begin{equation}
    \sket{\rho_i} = \sum_j \big(d(d+1) p^{(i)}_j - 1\big)\sket{\Pi_j}\mbox{.}
\end{equation}
Applying the Liouvillian to this expansion and using the eigenvalue equation, we obtain  
\begin{align*}
    &\vL\sket{\rho_i} = \sum_j \big(d(d+1) p^{(i)}_j - 1\big)\vL\sket{\Pi_j} = \lambda_i\sket{\rho_i} \\
    &\hspace{3cm}\Longrightarrow\\ 
    &\frac{d p_k^{(i)}}{dt} = \sum_j \big(d(d+1) p^{(i)}_j - 1\big)M_{k,j} = \lambda_i p_k^{(i)}\mbox{.}
\end{align*}
Thus we see that if $\lambda_i = 0$ the corresponding probability vector to the right eigenvector is a steady state of the SIC-POVM evolution.

It is also possible that one gets a periodic state in the time-asymptotic limit. This is easily seen by considering a Liouvillian whose spectrum has purely imaginary eigenvalues, that is
\begin{equation}
    \vL\sket{\rho_i} = i\lambda_i\sket{\rho_i}\mbox{.}
\end{equation}
Hence, a general asymptotic state is given by
\begin{align*}
    &\sket{\rho^{as}(t)} = \sum_{n}\rho_{n}(0)\sket{\rho_n^{(0)}} + \sum_{i} \rho_i(0)e^{i\lambda_i t}\sket{\rho_i^{+}} + \rho_i(0)^* e^{-i\lambda_i t} \sket{\rho_i^{-}}\mbox{,}
\end{align*}
whose corresponding probability vector in the SIC-POVM representation evolves in time since $(d/dt)\mbox{Tr}(\Pi_i\rho^{as}(t)) \ne 0$.

As an example, we examine a SIC-POVM representation of the evolution of a qubit undergoing spontaneous decay. The Liouvillian consists of the following Hamiltonian and jump operator
\begin{equation}
    \hat{H} = \frac{\Omega}{2}\hat{\sigma}_z + g\hat{\sigma}_x, \quad \hat{L} = \hat{\sigma}^-,
    \label{eq:qubitsicpovm_ev}
\end{equation}
where $\hat{\sigma}_{x,z}$ are the Pauli matrices ($\hat{\sigma}^-$ is the lowering operator), $\Omega$ is the energy separation between the states $\ket{0}$ and $\ket{1}$, and $g$ the coupling between them. Vectorizing the Liouvillian yields the matrix
\begin{equation}
    \vL = \begin{pmatrix}
            -\gamma & ig & -ig & 0 \\
            ig & (-\gamma/2 - i\Omega) & 0 & -ig \\
            -ig & 0 & (-\gamma/2 + i\Omega) & ig \\
            \gamma & -ig & ig & 0
        \end{pmatrix}.
        \label{eq:qubitL}
\end{equation}

To proceed, we choose a SIC-POVM set. In this case, we select the one represented by the states
\begin{align}
    &\ket{\psi_1} = \ket{0}, \quad \quad \ket{\psi_2} = \frac{1}{\sqrt{3}}\ket{0} + \sqrt{\frac{2}{3}}\ket{1}, \\
    &\ket{\psi_3} =\frac{1}{\sqrt{3}}\ket{0} + \sqrt{\frac{2}{3}}e^{i2\pi/3}\ket{1}, \quad \quad
    \ket{\psi_4} =\frac{1}{\sqrt{3}}\ket{0} + \sqrt{\frac{2}{3}}e^{i4\pi/3}\ket{1},
\end{align}
where the corresponding POVM elements are given by
\begin{equation}
    \Pi_i = \frac{1}{2} \ketbra{\psi_i}{\psi_i}, \quad i = 1,\,2,\,3,\,4.
\end{equation}
Using the matrix elements $\sbra{\Pi_i}\vL\sket{\Pi_j}$, we find, after some laborious calculations, that
\begin{equation}
    M = \begin{pmatrix}
            -\gamma/4 & -\gamma/12 & \frac{-\gamma + \sqrt{6}g}{12} & \frac{-\gamma - \sqrt{6}g}{12} \\
            \gamma/12 & -\gamma/36 & \frac{2\gamma - 2\sqrt{3}\Omega - \sqrt{6}g}{36} & \frac{2\gamma + 2\sqrt{3}\Omega + \sqrt{6}g}{36} \\
             \frac{\gamma - \sqrt{6}g}{12} & \frac{2\gamma + 2\sqrt{3}\Omega + \sqrt{6}g}{36} & -\gamma/36 & \frac{\gamma - \sqrt{3}\Omega + \sqrt{6}g}{18} \\
             \frac{\gamma + \sqrt{6}g}{12} & \frac{2\gamma - 2\sqrt{3}\Omega - \sqrt{6}g}{36} & \frac{\gamma + \sqrt{3}\Omega - \sqrt{6}g}{18} & -\gamma/36
        \end{pmatrix}\mbox{,}
        \label{eq:Mqubit}
\end{equation}
and that $\vec{m}$ is given by 
\begin{equation}
    \vec{m} = \begin{pmatrix}
        -\gamma/2, \gamma/6, \gamma/6, \gamma/6
    \end{pmatrix}^T.
    \label{eq:sicpovm_m}
\end{equation}
One can verify that $M$ satisfies the probability conservation condition, Eq.~(\ref{eq:consv_prob}).

\begin{figure}[!ht]
    \centering
    \includegraphics[width=0.4\linewidth]{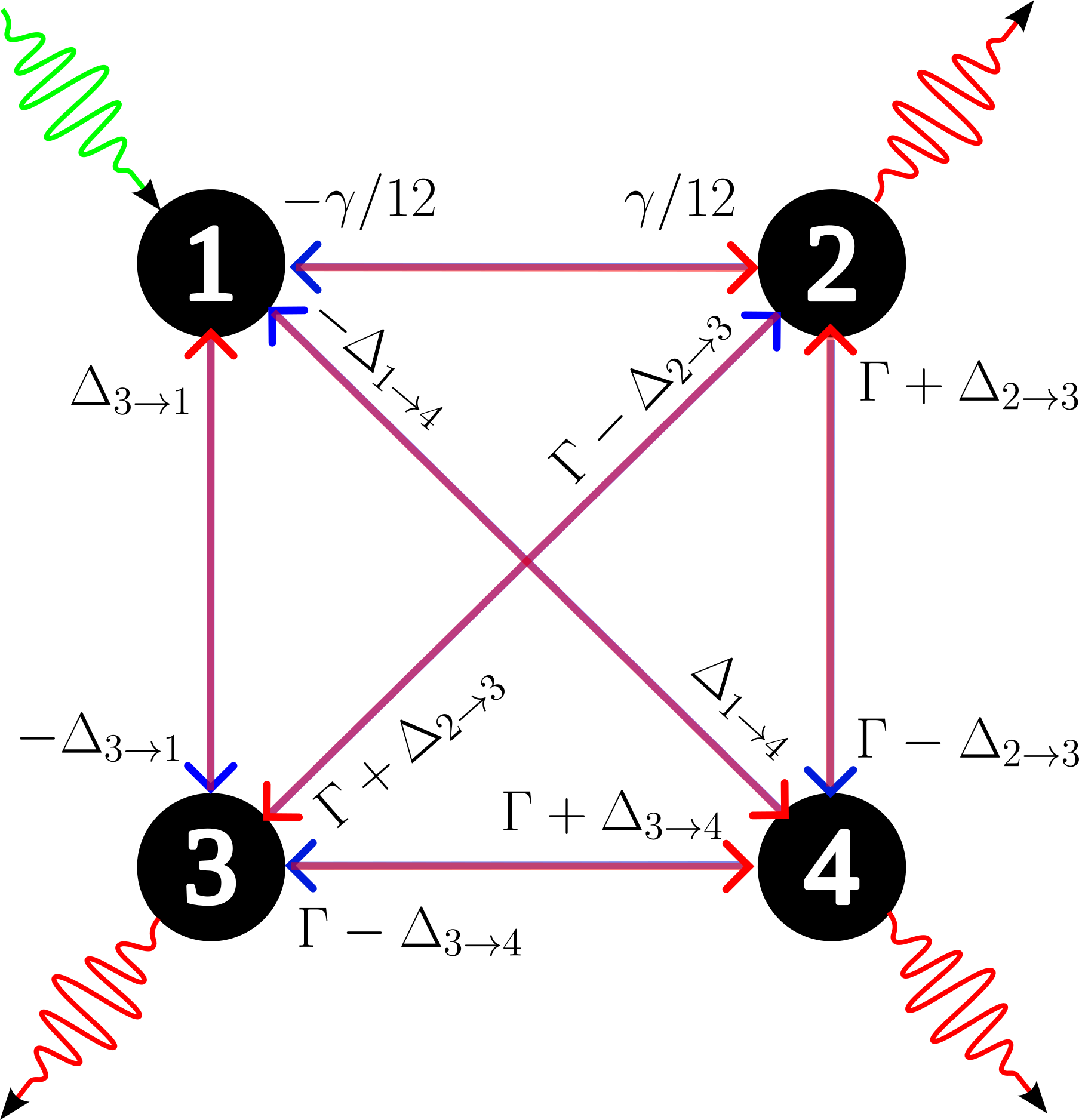}
    \caption{A schematic plot of the sites associated with the probabilities in the SIC-POVM representation of a qubit and the transitions between them, as given by the matrix $M$ Eq.~(\ref{eq:Mqubit}). These transitions are generated by the Lindblad open evolution with the Hamiltonian and jump operator given in Eq.~(\ref{eq:qubitsicpovm_ev}). The red photon-like arrow represents a constant input at site 1, while the green arrows indicate constant output/decay for the remaining sites. The red arrows illustrate positive transition amplitudes while blue represent negative ones. The transition amplitudes are given by: 
$\Gamma = \gamma/18$, 
$\Delta_{1\rightarrow4} = (\gamma + \sqrt{6})/12$, 
$\Delta_{3\rightarrow1} = (-\gamma + \sqrt{6}g)/12$, 
$\Delta_{2\rightarrow3} = (2\sqrt{3} + \sqrt{6}g)/36$, 
and $\Delta_{3\rightarrow 4} = \sqrt{3}\Omega - \sqrt{6}g$.}
    \label{fig:sic_povm_qubit}
\end{figure}

As the generator of evolution, we observe that, in addition to having negative diagonal entries—similar to classical stochastic matrices—negative off-diagonal elements can also appear, depending on the relative magnitudes of $\gamma$, $\Omega$, and $g$. This signifies a non-reciprocal interaction between the two states in the process, where one site can suppress transitions from other sites to itself. Furthermore, the components of $\vec{m}$ represent an on-site pump for site 1 and on-site losses for the other sites. In Fig.~\ref{fig:sic_povm_qubit}, we provide a graphical representation of this process.

To understand the consequences of having negative transition amplitudes, let us first analyze the case where $\gamma = 0$. Due to the non-commuting terms in the Hamiltonian, one expects periodic evolution rather than a steady state in the long-time limit. That is, the probabilities on each site oscillate around an average value — manifesting as Rabi oscillations in this model.

\begin{figure}[!ht]
    \centering
    \includegraphics[width=0.7\linewidth]{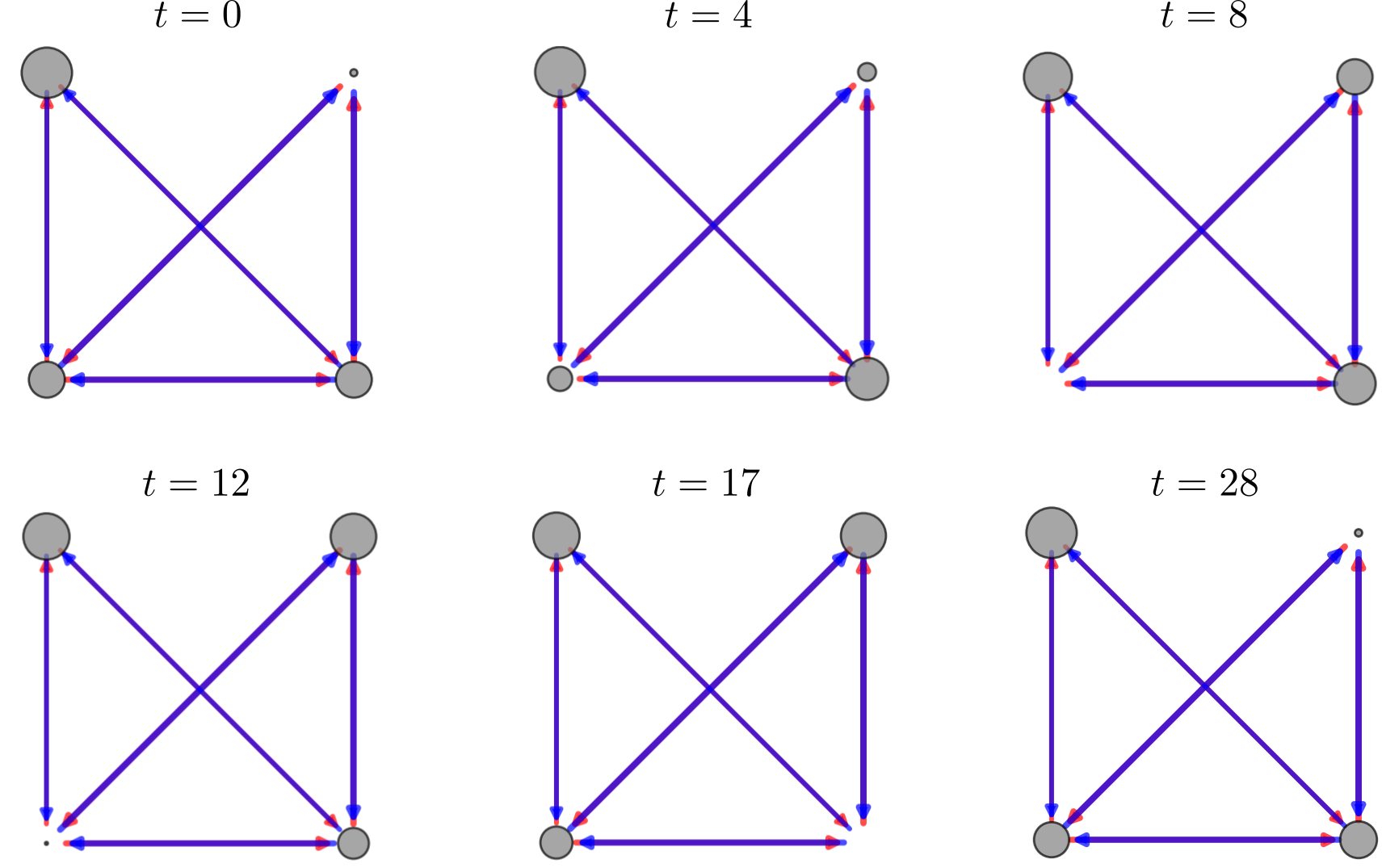}
    \caption{Snapshots of the evolution of the SIC-POVM probability vector, generated by the Liouvillian in Eq.~(\ref{eq:qubitL}), for $\gamma = 0$ with the initial state $\ket{0}$. The red directed arrows represent edges with positive transition amplitudes, while the blue edges indicate those with negative transition amplitudes. The size of each site is proportional to the probability of the system occupying the corresponding state.
}   
    \label{fig:sic_povm_unitary}
\end{figure}

First, we observe that the sites $1$ and $2$ are disconnected. Second, the negative transition rates 
$\Delta_{4\rightarrow1} = \Delta_{1\rightarrow3} = -(\sqrt{6}/12)g$ 
and $\Delta_{3 \rightarrow 2} = \Delta_{2 \rightarrow 4} = -(2\sqrt{3}\Omega + \sqrt{6}g)/36$ 
favor the cyclic path $1 \rightarrow 4 \rightarrow 2 \rightarrow 3 \rightarrow 1$ over the reverse direction. Furthermore, if $\Omega > \sqrt{2}g$, another cyclic path emerges: $2 \rightarrow 3 \rightarrow 4 \rightarrow 2$. Conversely, if $\Omega < \sqrt{2}g$, the preferred cycle becomes $1 \rightarrow 4 \rightarrow 3 \rightarrow 1$. We emphasize that the existence of a periodic state is a direct consequence of negative transition amplitudes. Figure~\ref{fig:sic_povm_unitary} presents snapshots of the probability evolution in the lattice.

If $\gamma \ne 0$, we expect the system to reach a steady state rather than exhibit periodic behavior. The steady-state distribution must satisfy Eq.~(\ref{eq:sicpovm_ss_cond}), meaning that the probability flow due to on-site terms is balanced by transitions between sites. Considering the case where $\Delta_{3 \rightarrow 1} < 0$, i.e., $\gamma > \sqrt{6}g$, the transition amplitude from site $3$ to site $1$ is suppressed. Consequently, the only positive probability flow into site $1$ comes from the on-site term. The balance equation then implies that, in general, a nonzero probability of occupying site $1$ — corresponding to the projector on state $\ket{0}$ — is achieved. This aligns with the intuition that the jump operator tends to drive the spin down, while the Hamiltonian term counteracts by driving the spin up. However, in the limit where $\gamma \gg g,\Omega$, dissipation dominates over coherent driving, effectively emptying state $1$. 

Using Eq.~(\ref{eq:sicpovm_ss_cond}) for $i = 1$ and approximating $-\gamma \pm \sqrt{6}g \approx -\gamma$, we obtain
\begin{equation}
    p_1^{\mbox{ss}} \approx \frac{\gamma/2 - \gamma/2(p_2^{\mbox{ss}} + p_3^{\mbox{ss}} + p_4^{\mbox{ss}}) }{\gamma/4} 
    \quad\leftrightarrow\quad p_1^{\mbox{ss}} \approx  2p_1^{\mbox{ss}}\quad \leftrightarrow\quad p_1^{\mbox{ss}} = 0\mbox{.}
\end{equation}
Following a similar reasoning, we can estimate the remaining steady-state probabilities in this regime. A similar calculation shows that $p_2^{\mbox{ss}} = p_3^{\mbox{ss}} = p_4^{\mbox{ss}} = 1/3$.

The steady-state solution can be found analytically from the steady-state density matrix, as mentioned earlier. To obtain the steady-state density matrix, we seek the null space of the Liouvillian matrix given in Eq.~(\ref{eq:qubitL}), namely  
\begin{equation}
    \vL \sket{\rho_{\mbox{ss}}}= \begin{pmatrix}
            -\gamma & ig & -ig & 0 \\
            ig & (-\gamma/2 - i\Omega) & 0 & -ig \\
            -ig & 0 & (-\gamma/2 + i\Omega) & ig \\
            \gamma & -ig & ig & 0
        \end{pmatrix}\begin{pmatrix}
                        a \\
                        b \\
                        c \\
                        d
                    \end{pmatrix} = 0\mbox{.}
\end{equation}
This yields the following system of equations
\begin{align*}
    -\gamma a + ig(b - c) = 0\mbox{,} \\
    ig\,a + (-\gamma/2 -i \Omega)b - ig\,d = 0\mbox{,} \\
    -ig\,a + (-\gamma/2 +i \Omega)c + ig\,d = 0\mbox{,} \\
    \gamma a - ig(b - c) = 0 \mbox{.}
\end{align*}
Solving this system and normalizing $\sket{\rho_{\mbox{ss}}}$ with respect to the vectorized identity, we obtain
\begin{equation}
    \sket{\rho_{\mbox{ss}}} = \frac{1}{2g^2 + \gamma^2/4 + \Omega^2}\begin{pmatrix}
                                                                        g^2 \\
                                                                        (-\Omega + i\gamma/2)g \\
                                                                        (-\Omega - i\gamma/2)g \\
                                                                        g^2 + \gamma^2/4 + \Omega^2
                                                                    \end{pmatrix}.
                                                                    \label{eq:sicpovm_ss}
\end{equation}
Notably, we confirm the expected result when $g = 0$, in which case the steady state simplifies to $\sket{\rho_{\mbox{ss}}} = \sket{1,1}$. Now, by using the definition of the SIC-POVM representation probabilities we obtain, $p_i^{\mbox{ss}} = \mbox{Tr}(\Pi_i\rho_{\mbox{ss}})$, and the steady state solution of the SIC-POVM evolution becomes
\begin{align}
    &p_1^{\mbox{ss}} = \frac{1}{2}\frac{g^2}{2g^2 + \gamma^2/4 + \Omega^2} \\
    &p_2^{\mbox{ss}} = \frac{1}{6}\frac{(3g^2 + \gamma^2 + 2\Omega^2 - 2\sqrt{2}\Omega g)}{2g^2 + \gamma^2/4 + \Omega^2} \\
    &p_3^{\mbox{ss}} = \frac{1}{6}\frac{(3g^2 + \gamma^2 + 2\Omega^2 +\sqrt{2}\Omega g - (\sqrt{6}/2)\gamma g)}{2g^2 + \gamma^2/4 + \Omega^2} \\
    &p_4^{\mbox{ss}} = \frac{1}{6}\frac{(3g^2 + \gamma^2 + 2\Omega^2 +\sqrt{2}\Omega g + (\sqrt{6}/2)\gamma g)}{2g^2 + \gamma^2/4 + \Omega^2}
\end{align}

Addressing how the relative magnitudes of the parameters affect the dynamics in the lattice when $\gamma \ne 0$, we first note that the cyclic path $2 \rightarrow 3 \rightarrow 4 \rightarrow 2$ might exist in the transient regime if $\gamma - (\sqrt{6}/2)g < \sqrt{3}\Omega$. Now, considering the case where $\Delta_{3\rightarrow 1} > 0$, the cyclic path $1 \rightarrow 4 \rightarrow 3 \rightarrow 1$ exists in the transient regime if  
\begin{equation}
    -\gamma + \sqrt{6}g < \sqrt{3}\Omega <  \gamma + \sqrt{6}g.
\end{equation}
The cyclic path that appears under closed evolution, $1 \rightarrow 4 \rightarrow 2 \rightarrow 3 \rightarrow 1$, exists if $\Delta_{2\rightarrow3} > \Gamma$, which translates to the condition  $\gamma < (\sqrt{6}/2)g$.
Finally, the cyclic path $2 \rightarrow 3 \rightarrow 4 \rightarrow 2$ is also possible if $\sqrt{3}\Omega > \gamma + \sqrt{6}g$.

\begin{figure}[!ht]
    \centering
    \includegraphics[width=0.7\linewidth]{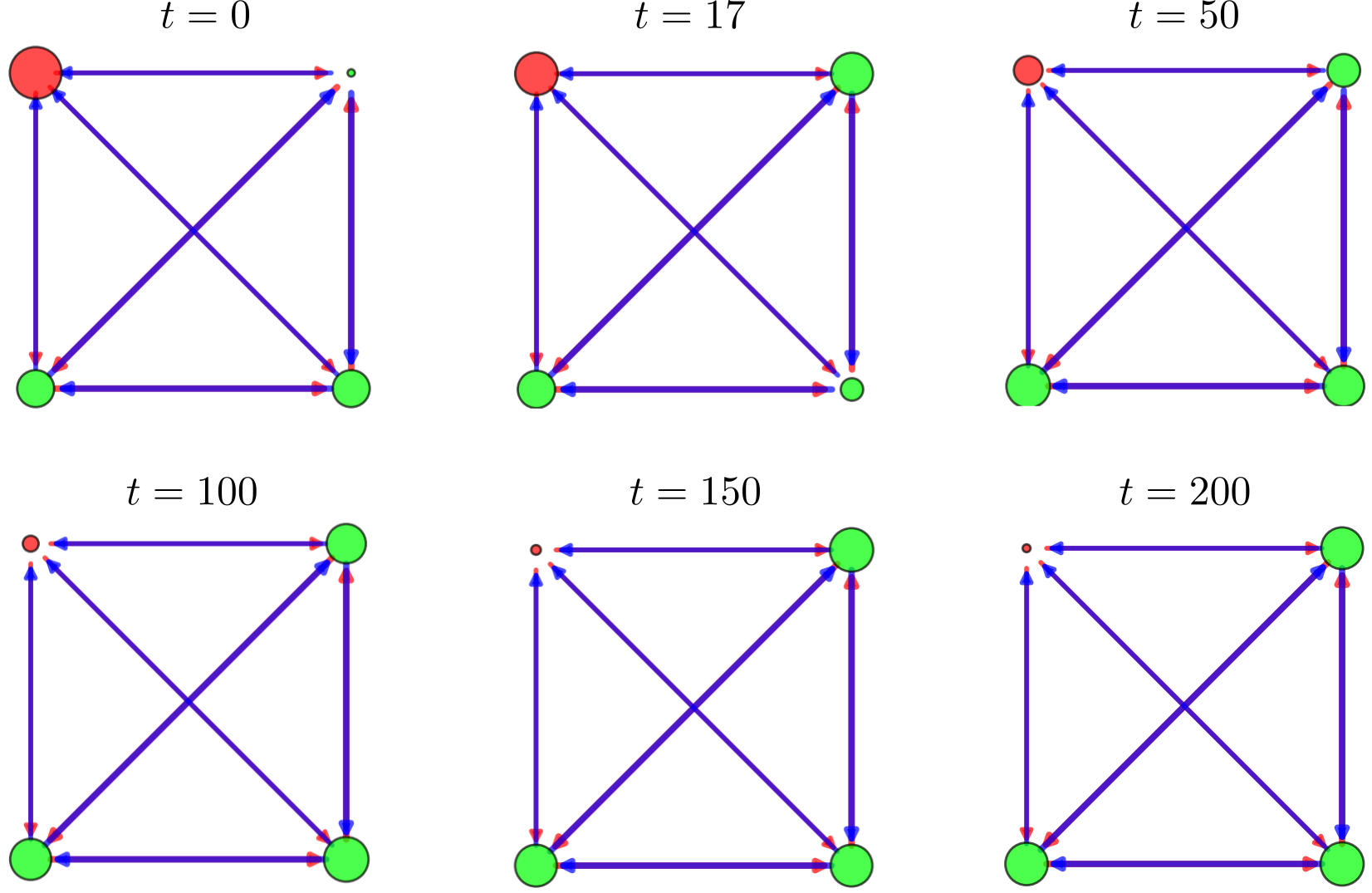}
    \caption{Snapshots of the evolution of the SIC-POVM probability vector generated by the Liouvillian Eq.~(\ref{eq:qubitL}) with $\Omega  = 1$, $g = 10^{-2}$, $\gamma  = 10^{-1}$, and initial state $\ket{0}$. The red directed arrows represent edges with positive transition amplitudes, while the blue edges represent those with negative ones. The size of the sites is proportional to the probability of the system being at their respective states. The red color represents an input of probability from the background, while the green color represents an output to the background, with the magnitudes given by the components of $\vec{m}$, as shown in Eq.~(\ref{eq:sicpovm_m}).
}
    \label{fig:sic_povm_open}
\end{figure}

In Fig.~\ref{fig:sic_povm_open}, we show the probability distribution evolution for some time steps when $\gamma > \sqrt{6}g$ and $\sqrt{3}\Omega > \gamma + \sqrt{6}g$.

\newpage 

\section*{Symmetries and Conserved Quantities in the LFSL} 
For Hamiltonian systems, the existence of a conserved quantity implies a continuous symmetry, and vice versa. In open systems, however, this correspondence does not always hold. Following~\cite{albert_symmetries_2014}, open systems exhibit two types of symmetries: \textit{weak symmetries} and \textit{strong symmetries}. 

As shown below, a weak symmetry is a symmetry of the Liouvillian that does not necessarily correspond to a conserved quantity. Let $\hat{S}$ be the Hermitian generator of a continuous symmetry of the Liouvillian, and define the unitary superoperator $\mathcal{U} = \exp(i\phi\mathcal{S})$, where $\mathcal{S}$ is the superoperator version of $\hat{S}$. A weak symmetry is then characterized by the relation  
\begin{equation}
    \mathcal{U}^\dagger \mathcal{L} \mathcal{U} = \mathcal{L}.
\end{equation} 
In contrast, a strong symmetry is one that is generated by an operator that commutes with every operator in the Liouvillian, i.e.,  
\begin{equation}\label{ssym}
    [\hat{S},\hat{H}] = [\hat{S}, \hat{L}_i] = 0.
\end{equation} 
A conserved quantity, in turn, is given by an operator $\hat{S}$ that satisfies $\partial_t\hat{S}=\mathcal{L}^\dagger\left[\hat{S}\right]=0$, or in vectorized form
\begin{equation}
    \frac{d}{dt}\sket{S} = \vL^{\dagger}\sket{S} = 0\mbox{,}
    \label{eq:conserved_qtd}
\end{equation} 
where we have used the definition of the adjoint of $\vL$ and transitioned to the Heisenberg picture. 

The presence of a weak symmetry can be formulated by the following identities~\cite{albert_symmetries_2014}
\begin{align}
    &[\hat{S}, \hat{H}] = 0 \label{eq:weak_cond1} \\
    &[\hat{S}, \hat{L}_k] = \alpha_k \hat{L}_k, \quad \alpha_k \in \mathbb{R}.
    \label{eq:weak_cond2}
\end{align}
This includes the special case of $\alpha_k=0$, which reproduces the condition~(\ref{ssym}) for a strong symmetry. It then follows that there is a conserved quantity related to the strong symmetry
\begin{equation}
    \mathcal{L}^{\dagger}(\hat{S}) = i[\hat{H}, \hat{S}] + \sum_k \frac{\gamma_k}{2}\big(2 \hat{L}_k \hat{S} \hat{L}_k^\dagger - \{\hat{L}_k^{\dagger}\hat{L}_k, \hat{S}\}\big) = 0.
    \label{eq:op_evol}
\end{equation}
We note that if $\alpha_k \neq 0$, we have a (weak) symmetry but not a conserved quantity since the last term in Eq.~(\ref{eq:op_evol}) is nonzero. Moreover, the existence of conserved quantities that are not related to any symmetry is also possible since Eq.~(\ref{eq:op_evol}) can vanish without $\hat{S}$ satisfying Eqs.~(\ref{eq:weak_cond1}) and ~(\ref{eq:weak_cond2}). 

To understand the role of symmetries and conserved quantities in LFSL dynamics, we need to clarify what a conserved quantity represents. The non-Hermiticity of $\vL$ implies that it may not be diagonalizable. In this case, its expansion must be performed in terms of generalized eigenvectors~\cite{gyamfi_fundamentals_2020}. However, assuming diagonalizability, the sets of left- and right-eigenvectors are generally distinct. Suppose there are $d$ linearly independent steady states, i.e., $d$ right-eigenvectors of $\vL$ with zero eigenvalues: $\vL\sket{\psi_i} = 0,\quad i = 1,\dots,d$. This implies that there exist at least $d$ linearly independent conserved quantities satisfying $\vL^\dagger\sket{J_i} = 0,\quad i = 1,\dots, d$. This is known as the \textit{steady-state conserved quantity correspondence}~\cite{albert_symmetries_2014}. 

Conserved quantities represent information retained from the initial state~\cite{fta} that persists throughout the dynamics. In the LFSL framework, this persistence manifests as non-vanishing populations. If $\sket{\rho_{in}}$ is the initial state and $\sket{\rho_{ss}} = \lim_{t \rightarrow \infty} e^{\vL t}\sket{\rho_{in}}$ is the steady state, then 
\begin{equation}
    \sbraket{J_i}{\rho_{ss}} = \lim_{t \rightarrow \infty}\langle\!\langle J_i\sket{e^{\vL t}\rho_{in}} = \lim_{t \rightarrow \infty}\sbra{e^{\vL^\dagger t}J_i}\rho_{in}\rangle\!\rangle = \sbraket{J_i}{\rho_{in}},
\end{equation}
where in the last equality we used the definition of a conserved quantity, Eq.~(\ref{eq:conserved_qtd}). 

In the LFSL, a continuous symmetry, as in the FSL, leads to a splitting of the lattice by reducing its dimension by one, while a discrete symmetry corresponds to breaking the lattice without reducing its dimension. However, in contrast to the FSL, here it is possible to have sublattices whose corresponding amplitudes vanish. These lattices represent decaying subspaces and emerge due to weak symmetries. Consequently, strong symmetries only partition the lattice into components containing non-vanishing amplitudes, as they necessarily imply conserved quantities. In the following section, we provide examples of LFSLs and analyze their associated symmetries and conserved quantities.

\newpage 

\section*{Examples of Liouville Fock-state Lattices}

\subsection{Open Jaynes-Cummings model}
Before introducing the LFSLs, we begin with a brief recapitulation of FSLs to build intuition on how LFSLs are constructed. To this end, we start by exploring the paradigmatic Jaynes-Cummings (JC) model, introduced in 1963 to describe coherent light-matter interaction~\cite{jaynes1963comparison,larson2021jaynes}. In this model, the atom is represented as a pseudo spin-1/2 particle, while only a single bosonic mode is considered. The total Hilbert space is given by $\mathfrak{H}_{JC} = \text{span} \{\ket{n,e(g)}, n \in \mathbb{N}\}$, where $\ket{n, e(g)} = \ket{n} \otimes \ket{e(g)}$. Here, $\ket{n}$ represents the state of the bosonic mode with $n$ excitations, while $\ket{e(g)}$ denotes the excited (ground) state of the atom. Note, when considering finite Hilbert spaces like spins, the Fock or number states are taken as those in which $\hat{\sigma}_z$ or $\hat{S}_z$ are diagonal. Hence, in the present example our Fock states are $\ket{n,e(g)}$. 

We begin with the \textit{quantum Rabi Hamiltonian}~\cite{rabi1937space,larson2021jaynes}, which describes the interaction of a quantized cavity mode with a two-level system
\begin{equation}
    \hat{H} = \hat{H}_{0} + \hat{H}_{\text{int}}, 
\end{equation}
where 
\begin{equation}
    \hat{H}_{0} = \omega \hat{n} + \frac{\Omega}{2}\hat{\sigma}_z, \hspace{1cm}
    \hat{H}_{\text{int}} = g(\hat{a} + \hat{a}^\dagger)\hat{\sigma}_x.
    \label{eq:quantum_rabi}
\end{equation}
Here, $g$ is the light-matter coupling strength, $\omega$ is the frequency of the bosonic mode, and $\Omega$ is the energy difference between the excited and ground states of the atom~\cite{ftb}. The number operator $\hat{n}$ gives the number of excitations in the bosonic mode, i.e., $\hat{n}\ket{n} = n\ket{n}$. The interaction term consists of a coupling between the creation and annihilation operators, whose action on the basis states follows the bosonic algebra
\begin{align}
    \hat{a}\ket{n} &= \sqrt{n}\ket{n-1}, \\
    \hat{a}^\dagger\ket{n} &= \sqrt{n+1}\ket{n+1}.
\end{align}
This is combined with the level-flip operator $\hat{\sigma}_x$, which satisfies $\hat{\sigma}_x\ket{e(g)} = \ket{g(e)}$. 

One can further decompose the interaction term into two parts: the \textit{rotating} and \textit{counter-rotating} terms
\begin{equation}
    \hat{H}_{\text{int}} = \hat{H}_{\text{rot}} + \hat{H}_{\text{crot}} 
    = g(\hat{\sigma}^+ \hat{a} + \hat{a}^\dagger\hat{\sigma}^-) 
    + g(\hat{a}\hat{\sigma}^- + \hat{a}^\dagger\hat{\sigma}^+).
\end{equation}
The rotating term describes the absorption (emission) of a photon accompanied by the excitation (decay) of the atom. In contrast, the counter-rotating term corresponds to simultaneous (de)excitation of both the atom and the field. Importantly, the latter term does not conserve the total excitation number operator
\begin{equation}
    \hat{N} = \hat{n} + \frac{1}{2}(\hat{\sigma}_z + 1).
    \label{eq:tot_exc}
\end{equation}
For sufficiently weak coupling, we can invoke the rotating-wave approximation (RWA)~\cite{larson2021jaynes}, which neglects the counter-rotating terms. This leads to the desired JC Hamiltonian
\begin{equation}
    \hat{H}_\mathrm{JC} = \frac{\Delta}{2}\hat{\sigma}_z + g(\hat{\sigma}^+ \hat{a} + \hat{a}^\dagger\hat{\sigma}^-),
    \label{eq:jc_model}
\end{equation}
where $\Delta = \Omega - \omega$ is the atom-field detuning.

To visualize the emerging FSL, we focus solely on the interaction term, as it is responsible for inducing transitions between the different basis states, which we take to represent our lattice sites. Under the RWA, only the transitions $\ket{n, g} \leftrightarrow \ket{n-1, e}$ are allowed, with coupling amplitude $g\sqrt{n}$. No other transitions between basis states occur. 

\begin{figure}[!ht]
    \centering
    \includegraphics[width=0.5\linewidth]{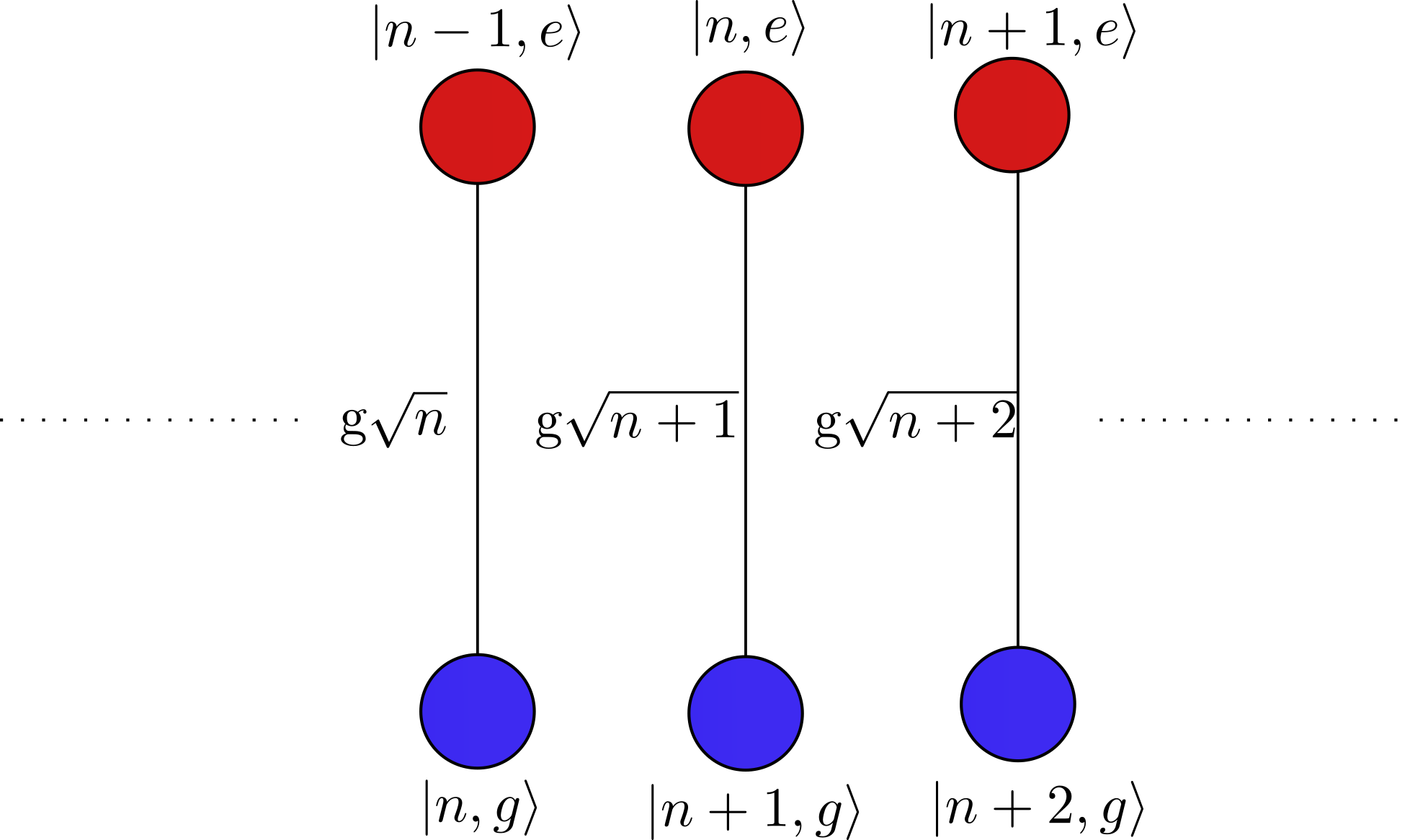}
    \caption{The FSL of the JC model, Eq.~(\ref{eq:jc_model}). The red circles represent sites where the atom is in the excited state, while the blue circles correspond to the atom in the ground state. The solid lines along the rungs indicate tunneling with rates $g\sqrt{n}$. Due to the assumed RWA, no transitions occur along the legs of the ladder. The single disconnected site, corresponding to the state $\ket{0,g}$, is not shown.}
    \label{fig:fsl_jc}
\end{figure}

As a result, the FSL of the JC model forms a ladder lattice, where there is no tunneling along the legs, but there is state-dependent tunneling along each rung. As such, the FSL forms an infinite set of two-site lattices (akin double-wells).  Additionally, note that a single site, $\ket{0,g}$, remains completely disconnected from the rest of the lattice. The FSL for the JC model is depicted in Fig.~\ref{fig:fsl_jc}.

\begin{figure}[!ht]
    \centering
    \includegraphics[width=0.5\linewidth]{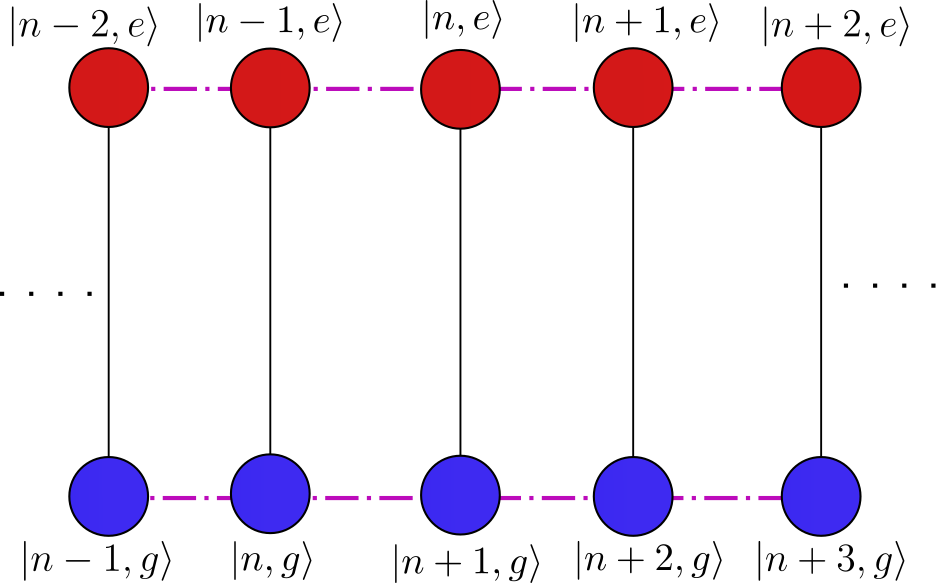}
    \caption{FSL of the driven JC model. Compared to the regular JC, the new tunnelings, denoted by purple dash-dotted lines, couple sites along the legs of the ladder. Note that the lattice is no longer separable, which derives from the fact that the drive breaks the continuous symmetry due to excitation conservation of the JC model.}
    \label{fig:fsl_djc}
\end{figure}

To the JC Hamiltonian, we can add a driving term,  
\begin{equation}
    \hat{H}_{\eta} = \eta(\hat{a} + \hat{a}^{\dagger}),
\end{equation}  
which couples states $\ket{n,e(g)} \leftrightarrow \ket{n\pm1, e(g)}$. This drive induces tunneling along the legs of the ladder in Fig.~\ref{fig:fsl_jc}, as illustrated in Fig.~\ref{fig:fsl_djc}.

\begin{figure}[!ht]
    \centering
    \includegraphics[width=0.5\linewidth]{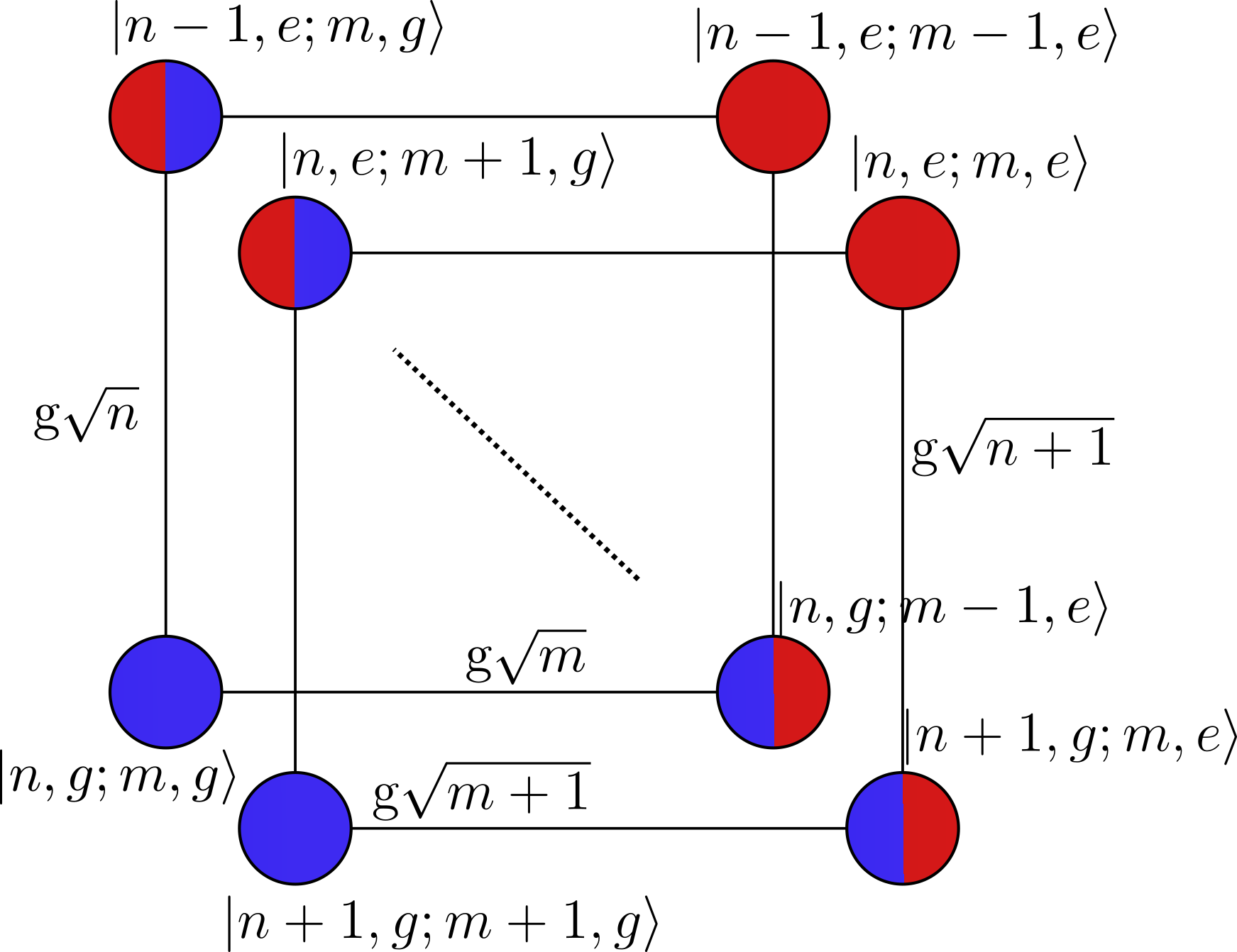}
    \caption{The LFSL of the JC model. To each site is attributed two colors, representing the possible elements of the atom's density matrix, $\sket{g,g}, \sket{g,e}, \sket{e,g}, \sket{e,e}$. The blue color represents the atom in the ground state, $\ket{g} \in \mathcal{H}$ on the left side and $\ket{g} \in \mathcal{H}^*$ on the right, while the red represents the excited state, $\ket{e} \in \mathcal{H}$ on the left side and $\ket{e} \in \mathcal{H}^*$ on the right. The transition amplitudes for each edge in the horizontal direction are the same within a given square, and the same applies to the edges in the vertical direction.}

    \label{fig:lfsl_jc}
\end{figure}

To take into account losses, we move to the Liouville space representation. Since in this representation density matrices are transformed into vectors, the set of states that span the space will be of the form 
\begin{equation}
    \sket{n,e; m,e}, \sket{n,e;m,g}, \sket{n,g;m,e}, \sket{n,g;m,g}, 
\end{equation}
with $n,m \in \mathbb{Z}_{+}$. Once the Liouvillian superoperator is given, the resulting LFSL readily follows. For the closed JC model Eq.~(\ref{eq:jc_model}), we find the vectorized Liouvillian
\begin{equation}
    \vL_\mathrm{JC} = -i\bigg(\frac{\Delta}{2}(\hat{\sigma}^z - \Tilde{\hat{\sigma}}^z) + g(\hat{a}^\dagger\hat{\sigma}^- - \Tilde{\hat{a}}^\dagger\Tilde{\hat{\sigma}}^-) + g(\hat{\sigma}^+\hat{a} - \Tilde{\hat{\sigma}}^+\Tilde{\hat{a}})\bigg)\mbox{,}
    \label{eq:liouville_jc}
\end{equation}
where, for notation simplicity, we omit writing out the tensor products with identities. Operators with a tilde belong to the doubled Hilbert space, $\Tilde{A} \in \mbox{Op}(\mathcal{H}^*)$.
It should be clear that the LFSL corresponding to the Liouvillian of Eq.~(\ref{eq:liouville_jc}) is given by an infinite set of four-site squares, see Fig.~\ref{fig:lfsl_jc}. Note that, however, not all lattices are squares. For instance, the state $\sket{0,g;0,g}$ is not connected to any other sites. In addition, states of the form $\sket{0,g;m,g}$ or $\sket{n,g;0,g}$ only form two-site lattices. As in Hilbert space, the LFSL of the JC model also lacks translational invariance.

Let us now introduce interactions with an external environment. First consider the JC model with a lossy atom. The simplest jump operator that we can consider is atom-decay $\hat{L} = \hat{\sigma}^-$. The Liouvillian becomes
\begin{equation}
    \vL = \vL_\mathrm{JC} + \frac{\gamma}{2}(2 \hat{\sigma}^- \otimes \Tilde{\hat{\sigma}}^- - \hat{\sigma}^+ \hat{\sigma}^- \otimes \mathbb{I} - \mathbb{I} \otimes \Tilde{\hat{\sigma}}^+\Tilde{\hat{\sigma}}^-)\mbox{,}
    \label{eq:openb_jc}
\end{equation}
with $\vL_{JC}$ given by Eq.~(\ref{eq:liouville_jc}). 

\begin{figure}[!ht]
    \centering
    \includegraphics[width=0.9\linewidth]{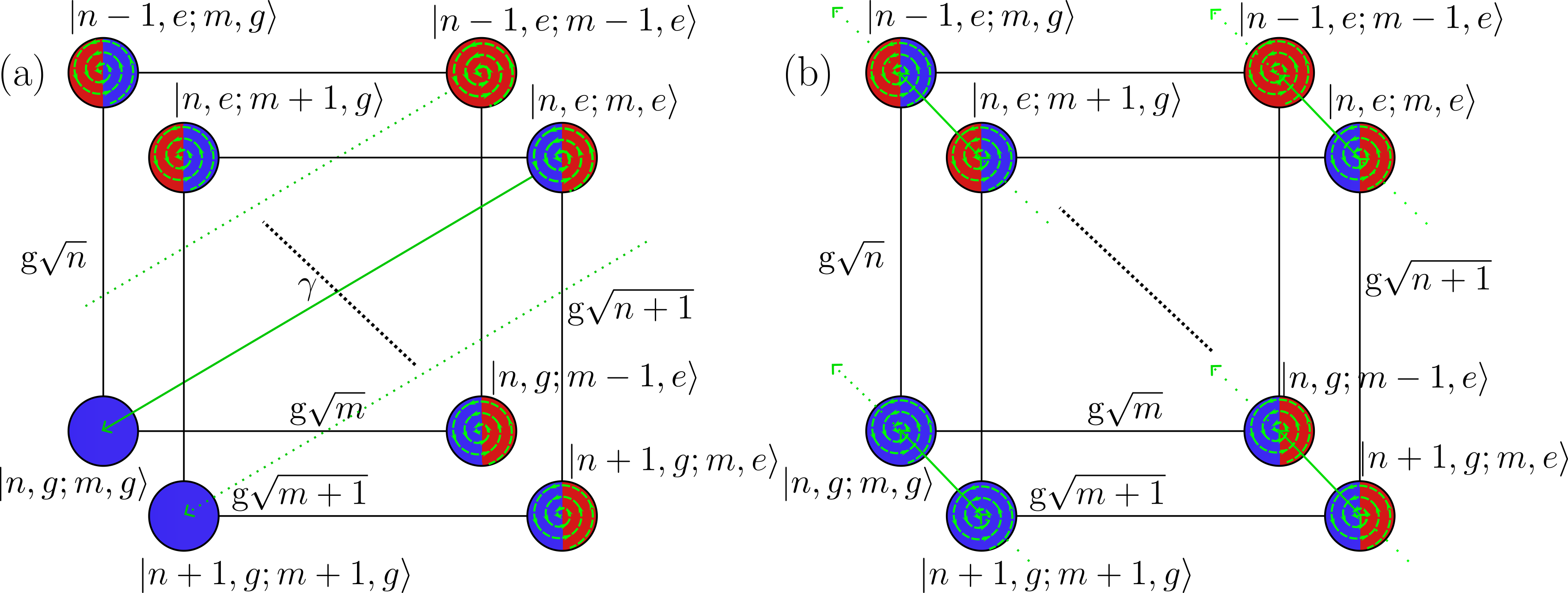}
  \caption{LFSLs for the open JC model with atomic decay $\hat{L} = \hat{\sigma}^-$ (a) and single-photon loss $\hat{L} = \hat{a}$ (b). The green arrows represent state transitions induced by the jump operator, while the green spirals indicate onsite amplitude decay. Hence, at sites marked with a green spiral, the population decreases without tunneling to another site.}

    \label{fig:lfsl_ojc}
\end{figure}

By applying the Liouvillian to the basis states, we see again that the part corresponding to the unitary dynamics gives us squares as in Fig.~\ref{fig:lfsl_jc}. Nonetheless, the jump terms now are going to couple each square. In Fig.~\ref{fig:lfsl_ojc} (a) we depict the LFSL of this model. Without losses, we had that the excitation number gives us a conserved quantity, $[\hat{N},\hat{H}_\mathrm{JC}] = 0$, and consequently a continuous symmetry of the lattice. To check if this is still the case in the presence of atomic decay, we need to verify the condition given by Eq.~(\ref{eq:weak_cond2}), i.e.,  
\begin{align}
    &[\hat{N} , \hat{L}] = \left[ \frac{1}{2}(\hat{\sigma}^z + \mathbb{I}) + \hat{a}^\dagger \hat{a}, \hat{\sigma}^- \right]  = \frac{1}{2}[\hat{\sigma}^z, \hat{\sigma}^-] = -\hat{\sigma}^-\mbox{.}
\end{align}
Therefore, Eq.~(\ref{eq:weak_cond2}) is satisfied with $\alpha = -1$, giving a weak symmetry instead of a strong one -- as we could have guessed since the number of excitations is obviously not preserved.  

As previously discussed, weak symmetries break up the lattice into sublattices with, in general, decaying amplitudes at the sites. Spontaneous emission described by $\hat{L}=\hat{\sigma}^-$ couples states $\sket{e, n; e, m}$ to states $\sket{g, n; g, m}$. We note that the difference between the total number of excitations for each space $\mathfrak{H}$ and $\mathfrak{H}^*$ remains constant despite this decay, i.e., $\hat{N} - \hat{\Tilde{N}} = C$ for some integer constant $C$.

 This model does not support any strong symmetry which might indicate that we do not have conserved quantities. In this sense, the LFSL is of great help in visualizing if that is indeed the case. Looking to Fig.~\ref{fig:lfsl_ojc}(a) we see that states of the form $\sket{n,g;m,g}$ are dark states, i.e. states that are invisible to the action of jump operators. For them to correspond to eigenstates of the Liouvillian they must also be eigenstates of the Hamiltonian. For a given set of eigenstates of the Hamiltonian that are also dark states $\{\ket{\psi_n}\}$, with corresponding eigenvalues $\lambda_n$, they form eigenstates of the Liouvillian $\sket{\rho_{n,m}} = \ket{\psi_n} \otimes \ket{\psi_m}^*$ 
 
 \begin{equation}
    \vL\sket{\rho_{n,m}} = -i(\lambda_n - \lambda_m)\sket{\rho_{n,m}}\mbox{ .}
\end{equation}
In this model not all of the identified dark states are eigenstates of the Hamiltonian and neither linear combinations of them can be. One can see that by noting that they are not connected to each other through the Hamiltonian interaction. However, there is one state that is an eigenstate of the Hamiltonian and a dark state at the same time which is the state $\sket{0,g;0,g}$. This state is the steady state of our model and this can be seen from the LFSL since it is the unique state that cannot transition to a decaying state due to the directional coupling.

Another option is to introduce single-photon loss, i.e., $\hat{L} = \hat{a}$. The corresponding Liouvillian is given by
\begin{equation}
    \vL = \vL_{JC} + \frac{\gamma}{2} \left( 2 \hat{a} \otimes \hat{\Tilde{a}} - \hat{a}^\dagger \hat{a} \otimes \mathbb{I} - \mathbb{I} \otimes \hat{\Tilde{a}}^\dagger \hat{\Tilde{a}} \right)\mbox{.}
    \label{eq:openb_jc}
\end{equation}
Once again, this model exhibits a weak symmetry generated by the total excitation number operator, as shown by
\begin{align}
    &[\hat{N} , \hat{L}] = \left[ \frac{1}{2}(\hat{\sigma}^z + \mathbb{I}) + \hat{a}^\dagger \hat{a}, \hat{a} \right]  = [\hat{a}^\dagger \hat{a}, \hat{a}] = -\hat{a}\mbox{.}
    \label{eq:weaksym_jcbloss}
\end{align}
Similar to the case of atomic decay, this weak symmetry results in a partitioning of the lattice such that the difference between the total excitation number in $\mathfrak{H}$ and $\mathfrak{H}^*$ remains constant. However, in this case, the squares are connected through all four types of states since the jump operator acts only on the bosonic mode, as illustrated in Fig.~\ref{fig:lfsl_ojc}(b). 

Additionally, it follows that the dark states, i.e., states that are unaffected by losses, are those in which no bosons are present, i.e., $\sket{0, e(g); 0, e(g)}$. Since $\sket{0,g;0,g}$ is the only such state that is also an eigenstate of the Hamiltonian, it follows that this is the unique steady state~\cite{larson2017some}.

\subsection*{Lossy quantum Rabi model}
\begin{figure}[!ht]
    \centering
    \includegraphics[width=0.5\linewidth]{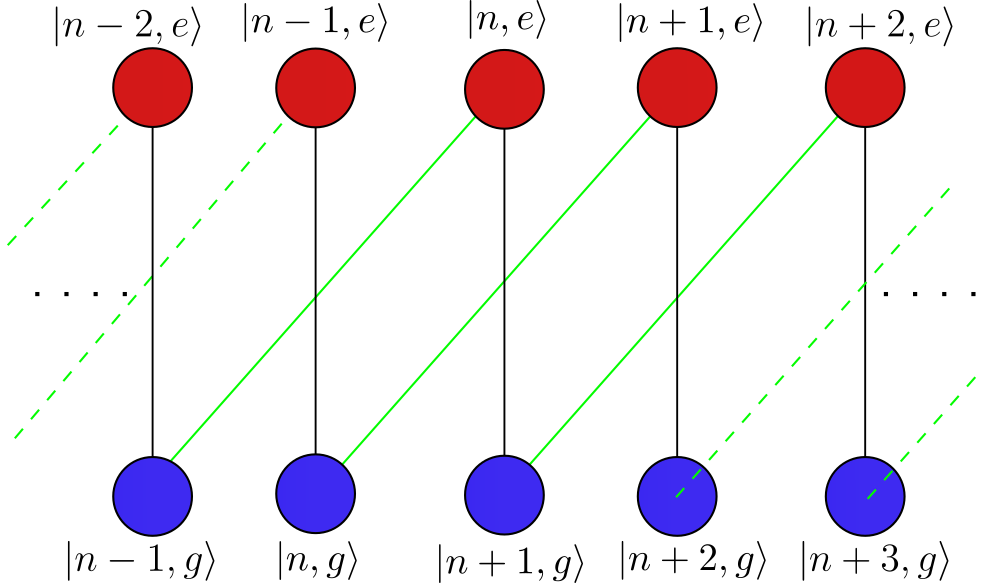}
    \caption{FSL of the quantum Rabi model. The transitions induced by the counter-rotating term, which is present in the quantum Rabi model but not in the JC model, are represented by the green lines. Note how the ladder structure splits into two chains, one for each parity of the quantum Rabi model. As a result, the FSL can be redrawn as two disconnected 1D lattices.}
    \label{fig:fsl_qrm}
\end{figure}

By including the counter-rotating term, i.e., the full interaction, we derive the quantum Rabi model~(\ref{eq:quantum_rabi}) rather than the JC model. In this case, the lattice consists of two one-dimensional lattices, each corresponding to a parity sector of the total excitation number. This is because, in addition to the transitions that preserve the excitation number, we also have $\ket{n,e} \leftrightarrow \ket{n-1,g}$, which increases/decreases the total number of excitations in pairs. The FSL for the quantum Rabi model can be seen in Fig.~\ref{fig:fsl_qrm}.

A slightly more general case is the \textit{anisotropic quantum Rabi model}~\cite{xie2014anisotropic}, where we consider different couplings for the rotating and counter-rotating terms:
\begin{equation}
    H_{\mbox{\tiny int}}^{\mbox{\tiny ani}} = g_c(a\sigma^+ + a^\dagger\sigma^-) + g_{cr}(a^\dagger\sigma^+ + a\sigma^-)\mbox{ .}
\end{equation}
The anisotropic nature of the model, $g_c \neq g_{cr}$, does not alter the FSL geometry. However, due to the alternating tunneling amplitudes resulting from this, the FSL becomes similar to an SSH chain with topological edge states~\cite{saugmann2023fock}.

\begin{figure}
    \centering
    \includegraphics[width=0.9\linewidth]{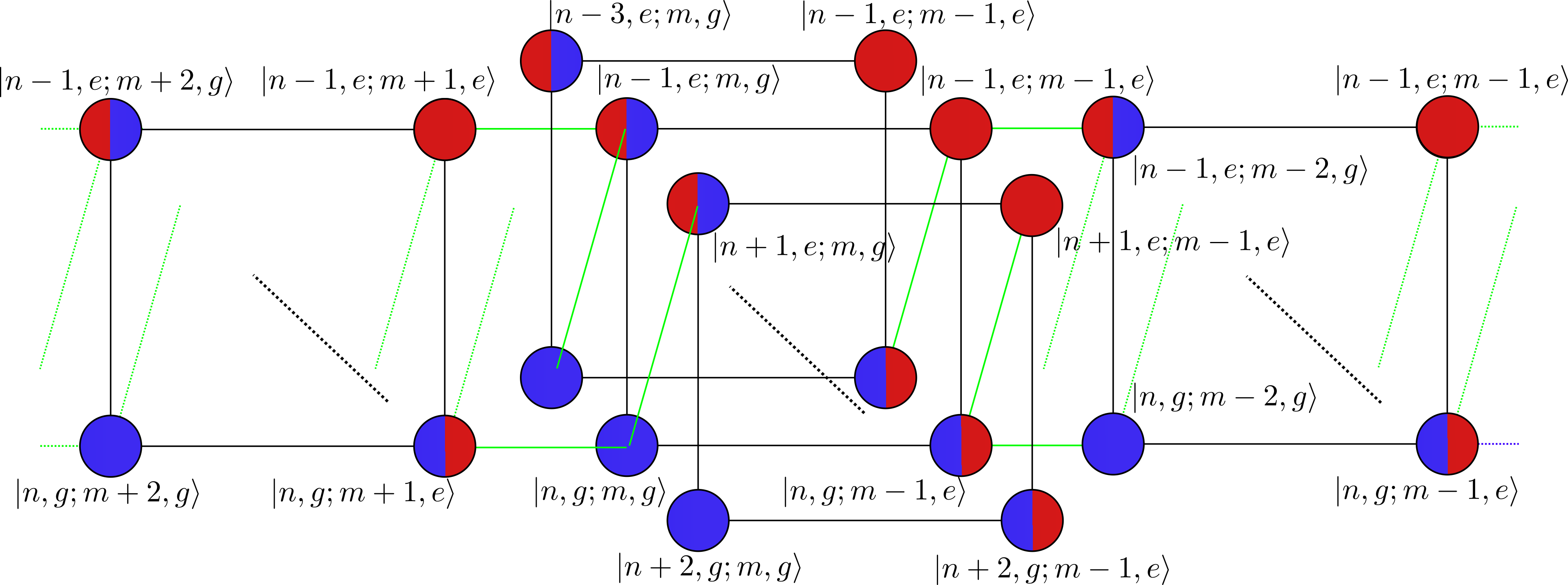}
    \caption{LFSL of the quantum Rabi model without any losses. Like for the JC model, red and blue sites represent the internal atomic states $|e\rangle$ and $|b\rangle$ respectively, while the blue/red dots are the coherences. Black lines are the edges corresponding to transitions induced by the rotating terms while the green edges are induced by the counter-rotating terms. }
    \label{fig:LFSL_QRM}
\end{figure}

The Liouvillian of the quantum Rabi model is given by, ignoring the free terms,
\begin{equation}
    \vL_{QR} = -ig\left[(\hat{a} + \hat{a}^\dagger)(\hat{\sigma}^+ + \hat{\sigma}^-) - (\hat{\Tilde{a}} + \hat{\Tilde{a}}^\dagger)(\Tilde{\sigma}^+ + \Tilde{\sigma}^-)\right]\mbox{.}
\end{equation}
Previously, we saw that the LFSL of the JC model is an infinite set of coupled square lattices, a consequence of the conservation of the total number of excitations. In the quantum Rabi model, these lattices become connected due to the counter-rotating term. On the other hand, due to the preservation of the parity of $N$ and $\Tilde{N}$, the total LFSL will consist of four rectangular parallelepipeds, as depicted in Fig.~\ref{fig:LFSL_QRM}.

By introducing atom decay, i.e. $\hat{L} = \hat{\sigma}^-$, the four possible rectangular parallelepipeds become connected to each other due to the change of the total excitation number's parity. We do not try to draw it here since a depiction becomes quite messy. The same would happen if we introduce one-photon decay.

\subsection*{Incoherently driven Oscillator}
Another model that we present here is the incoherently driven oscillator, i.e. a quantum oscillator that besides being driven coherently by a Hamiltonian term it is also under the action of a drive jump operator, i.e.
\begin{align}\label{indrive}
    \hat{H} &= \Delta\hat{a}^\dagger\hat{a}+\eta\left(\hat{a}^\dagger+\hat{a}\right)\mbox{ ,}\\
     \hat{L} &= \hat{a}^\dagger+\hat{a}\mbox{ .}
\end{align}

\begin{figure}[!ht]
    \centering
    \includegraphics[width=0.5\linewidth]{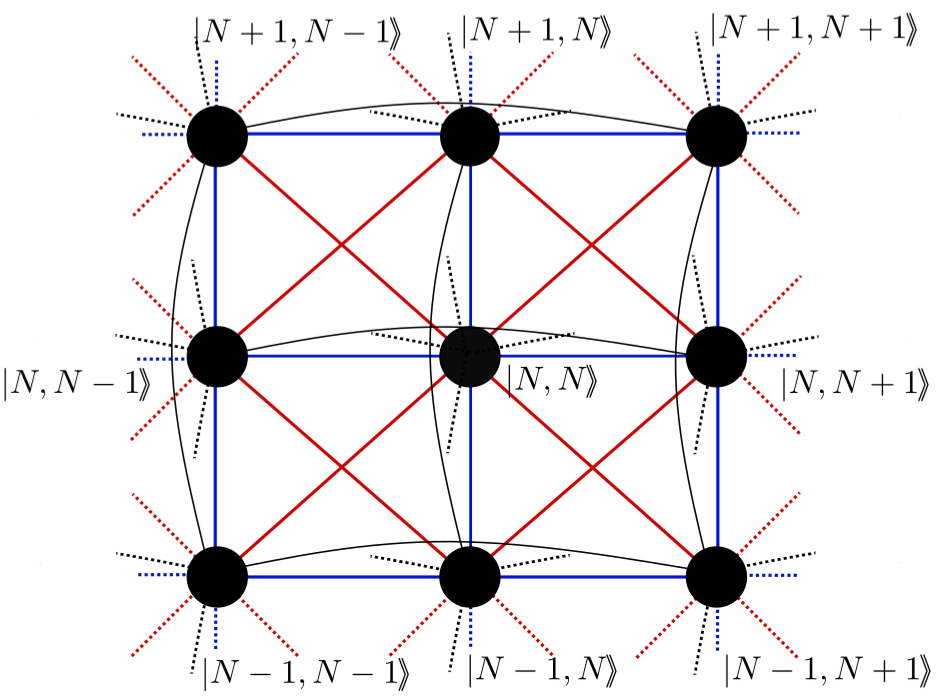}
    \caption{LFSL of the incoherently driven oscillator, Eq.~(\ref{indrive}). Here, we identify three types of tunneling: solid blue lines represent those arising from coherent driving (Hamiltonian term $\eta\left(\hat{a}+\hat{a}^\dagger\right)$); solid red lines correspond to tunneling induced by $\hat{L}\hat{\Tilde{L}}^\dagger$ and terms like $\hat{a}^\dagger\hat{\Tilde{a}}$; finally, solid black lines represent tunneling due to $(\hat{a}^\dagger)^2$ and $\hat{a}^2$, which stem from $\hat{L}^\dagger\hat{L}$ (and similarly for $\hat{\Tilde{L}}^\dagger\hat{\Tilde{L}}$).}

    \label{fig:incoherent_driven_oscillator}
\end{figure}

Since the FSL of the driven oscillator is a one-dimensional chain, the LFSL forms a square lattice. As in the case of a one-photon gain or loss, i.e., $\hat{L}=\hat{a}^\dagger$ or $\hat{L}=\hat{a}$, the incoherent drive induces diagonal hopping in the square lattice, arising from terms like $\hat{a}^\dagger\hat{\Tilde{a}}$. Consequently, the lattice is not simply a tight-binding, or nearest-neighbor, lattice. Moreover, the terms in the Liouvillian resulting from $\hat{L}^\dagger \hat{L}$ introduce next-next-nearest-neighbor hopping due to $(\hat{a}^\dagger)^2$ and $\hat{a}^2$. We show the LFSL of this model, along with its different tunneling terms, in Fig.~\ref{fig:incoherent_driven_oscillator}.

This model is somewhat special since there is no biased flow in the LFSL; the tunneling rates are equally strong in both directions. Nevertheless, it is possible to find the steady state analytically by noting that we can write $\hat{L} = \sqrt{2}\hat{x}$ in the quadrature representation of the boson operators, i.e., $\hat{x}=\frac{1}{\sqrt{2}}\left(\hat{a}+\hat{a}^\dagger\right)$ and $\hat{p}=\frac{i}{\sqrt{2}}\left(\hat{a}-\hat{a}^\dagger\right)$. The jump operator has the effect of localizing the particle in the $x$-quadrature representation. 
Recognizing this, the most general steady state can be expressed as  
\begin{equation}
\hat{\rho}^\mathrm{ss}=\sum_n p_n \hat{\rho}_n^\mathrm{ss}, \hspace{1cm} \text{with} \hspace{1cm} \hat{\rho}_n^\mathrm{ss}=\int dx\,|\psi_n(x)|^2 |x\rangle\langle x|\mbox{ ,}
\end{equation}  
where $\psi_n(x)$ are the eigenfunctions of the Hamiltonian, and the probabilities $p_n$ depend on the system's initial state.

To find the spectrum of the Liouvillian, we resort to \textit{third quantization}~\cite{prosen2008third}, a method for solving quadratic Liouvillians. Originally applied to fermionic Lindbladians, it has recently been extended to the bosonic case~\cite{prosen_quantization_2010, kim_third_2023}. The resulting spectrum is purely imaginary and equidistant, with the spacing determined by the effective frequency $\Delta$, i.e.  
\begin{equation}
    \mu_{n,m} = i(n - m)\Delta\mbox{, }\hspace{1cm}n,m \in \mathbb{N}\mbox{.}
\end{equation}  
Note that since $n$ and $m$ are natural numbers, the system possesses an infinite number of conserved quantities.  

In the FSL of the Hamiltonian, the term $\Delta\hat{a}^\dagger\hat{a}$ introduces a linear onsite energy offset, while the drive induces nearest-neighbor tunneling in the FSL, reminiscent of the Wannier-Stark problem. Similarly, in the Wannier-Stark Hamiltonian, the spectrum is also independent of the tunneling rate, while the eigenstates remain dependent on it~\cite{hartmann2004dynamics}.

\subsection*{Central spin model}
In the \textit{central spin model}, a single spin-$1/2$ particle interacts identically with $N$ spin-$1/2$ particles~\cite{mermin1991can,breuer2004non,larson2021jaynes}. By utilizing the conservation of total spin, $S \leq N/2$, we introduce the collective spin operators $\hat{S}_\alpha=\sum_{j=1}^N\hat{\sigma}_\alpha^{(j)}$, with $\alpha = x, y, z$, where $\hat{\sigma}_\alpha^{(j)}$ are the Pauli matrices for spin $j$. The Hamiltonian can then be written as  
\begin{equation}
    \hat{H}_\mathrm{cs}=\delta\hat{\sigma}_z+\Delta\hat{S}_z+g\hat{\sigma}_x\hat{S}_x.
\end{equation}  
Here, $\delta$ and $\Delta$ are external "fields" that attempt to align the spins in the $z$-direction, while $g$ is the coupling strength, as before. The Fock states in this model are given by $\ket{e(g),m}$, where $\hat{\sigma}_z\ket{e(g),m}=\pm\ket{e(g),m}$ (with $+$ for $e$ and $-$ for $g$) and $\hat{S}_z\ket{e(g),m}=m\ket{e(g),m}$ for $-S\leq m\leq S$. Additionally, we have $\hat{S}_x=(\hat{S}^++\hat{S}^-)/2$, with the ladder operators acting as  
\begin{equation}
    \hat{S}^\pm\ket{S,m}=\sqrt{(S\mp m)(S\pm m+1)}\ket{S,m\pm1}.
\end{equation}  

This model can be viewed as a "poor man's" version of either the quantum Rabi or Dicke model~\cite{larson2021jaynes,mumford2014impurity}. Specifically, replacing either the large spin or the spin-$1/2$ particle with a bosonic mode, i.e., letting $\hat{\sigma}^\pm\rightarrow\hat{a},\,\hat{a}^\dagger$ or $\hat{S}^\pm\rightarrow\hat{a},\,\hat{a}^\dagger$, reproduces the quantum Rabi or Dicke models, respectively.  

The poor man's models have finite FSLs, and in the case of the central spin model, the FSL consists of two decoupled one-dimensional chains corresponding to each parity symmetry. Consequently, the LFSL forms a square lattice similar to that of the quantum Rabi model but with a finite size.

\subsection*{Optical bistability model}
Optical bistability occurs in driven/lossy cavity systems, where the cavity contains a nonlinear medium of some sort. Upon varying the drive frequency, the cavity field amplitude may exhibit multistability, manifesting as sudden jumps in the field intensity. In the bad cavity limit, when adiabatic elimination of the cavity field is justified, and assuming that the nonlinear medium consists of a set of identical spin-$1/2$ particles, one can derive a particularly simple model where the steady state is analytically known~\cite{bonifacio1978optical,bonifacio1978photon,agrawal1979optical,drummond1980quantum,rodriguez2017probing,hannukainen2018dissipation}. The Liouvillian takes the form  
\begin{equation}
    \mathcal{L}(\hat{\rho})=-i\left[\hat{\rho},\omega\hat{S}_x\right]+\frac{\gamma}{2}\left(2\hat{S}^-\hat{\rho}\hat{S}^+-\hat{S}^+\hat{S}^-\hat{\rho}-\hat{\rho}\hat{S}^+\hat{S}^-\right),
\end{equation}
or in its vectorized form,
\begin{equation}
    \bar{\mathcal{L}} = i\omega \left( \hat{S}_x - \hat{T}_x\right) 
    + \frac{\gamma}{2} \left( 2 \hat{S}^+ \hat{T}^- - \hat{S}^+\hat{S}^- - \hat{T}^+\hat{T}^- \right).
\end{equation}
Here, $\omega$ is an effective drive frequency of the spins, and as before, $\hat{S}_\alpha$ are the collective spin operators. We have omitted the explicit tensor product with identity operators, where the $\hat{S}$-operators act on the Hilbert space $\mathfrak{H}$, while the $\hat{T}$-operators act on the doubled Hilbert space $\mathfrak{H}^*$.  

This model can be viewed as a "poor man's" version of a lossy, resonantly driven oscillator with Hamiltonian $\hat{H}=\eta\left(\hat{a}+\hat{a}^\dagger\right)$ and jump operator $\hat{L}=\hat{a}$. Both models share the same type of LFSL -- a square lattice with nearest-neighbor tunneling rates stemming from the coherent pump and directional diagonal tunneling rates due to Lindblad terms, with the key difference that the present model has a finite dimension. This distinction qualitatively alters the physics of the two models, leading to critical behavior in the bistable model but not in the oscillator model. The oscillator has a coherent steady state,  
\begin{equation}
    \hat{\rho}_\mathrm{ss}=|\alpha\rangle\langle\alpha|,
\end{equation}  
with $\alpha=2i\eta/\gamma$. In contrast, the steady state for the bistability model can be expressed as~\cite{bonifacio1978optical,bonifacio1978photon}  
\begin{equation}
    \hat{\rho}_\mathrm{ss}=\hat{\chi}\hat{\chi}^\dagger,
\end{equation}
where  
\begin{equation}
    \hat{\chi}=\frac{1}{\sqrt{D}}\sum_{n=0}^{2S}\left(\frac{\hat{S}^-}{\lambda}\right)^n,\hspace{0.6cm}
    \mathrm{with}\hspace{0.6cm}
    D=\sum_{m=0}^{2S}\frac{(2S+m+1)!(m!)^2}{(2S-m)!(2m+1)!}|\lambda|^{-2m},\hspace{0.5cm}
    \mathrm{and}\hspace{0.5cm}
    \lambda=-i\omega S/\gamma.
\end{equation}
Upon closer inspection, the steady state exhibits a nonequilibrium phase transition at $\omega=\gamma$ in the thermodynamic limit $S\rightarrow\infty$~\cite{hannukainen2018dissipation}. For $\omega<\gamma$, the system becomes magnetized with $\langle\hat{S}_z\rangle\neq0$, whereas for $\omega>\gamma$, it transitions to a paramagnetic (or delocalized) phase with $\langle\hat{S}_z\rangle=0$. These phases are fundamentally different from the coherent steady state of the driven lossy oscillator. Hence, while both models share the same LFSL structure, the presence of an "upper" boundary in the bistability model leads to qualitatively different steady states.

\subsection*{Open tight-binding model}
Here, we present a more thorough analysis of the open ladder model discussed in the main text. The model is defined by  
\begin{align}\label{1dlat}
  \hat{H} &= \eta\sum_{n = -\infty}^\infty (\ketbra{n+1}{n} + \ketbra{n}{n+1}), \\ \nonumber \\
  \hat{L}_1 &= \sum_n \ketbra{n}{n+1}, \hspace{0.5cm} \hat{L}_2  = \sum_n \ketbra{n+1}{n}\nonumber.
\end{align}  
To solve this model, we resort to the vectorized representation of the Liouville superoperator. Since the model is translationally invariant, we consider it in quasi-momentum space. The quasi-momentum states  
\begin{equation}
    \ket{\theta} = \sum_n e^{i\theta n}\ket{n}, \hspace{0.5cm} \mbox{with} \hspace{0.5cm} \theta \in \left[-\frac{\pi}{2}, \frac{\pi}{2}\right)
\end{equation}
are eigenstates of the left and right shift operators $\hat{L}_1$ and $\hat{L}_2$ with eigenvalues $e^{\pm i\theta}$:  
\begin{align}
    \hat{L}_1\ket{\theta} &= e^{i\theta}\ket{\theta}, \\
    \hat{L}_2\ket{\theta} &= e^{-i\theta}\ket{\theta}.
\end{align}  
Note that in the extended space, i.e., $\mathfrak{H}^*$, the corresponding quasi-momentum states and the action of the left and right shift operators differ from the above by complex conjugation.

\begin{figure}[!ht]
    \centering
    \includegraphics[width=0.5\linewidth]{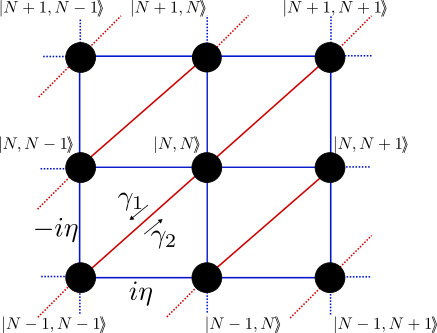}
    \caption{LFSL of the open one-dimensional lattice model with coherent and incoherent nearest-neighbor tunneling~(\ref{1dlat}). In the LFSL, coherent tunneling manifests as nearest-neighbor hopping (blue lines), while incoherent tunneling results in directional diagonal next-nearest-neighbor hopping (red lines). Note that when $\gamma_1 = \gamma_2$, the diagonal tunneling terms are balanced.}

    \label{fig:lattice_model}
\end{figure}

The action of the vectorized Liouville superoperator on the momentum states $\sket{\theta, \Tilde{\theta}}$ is given by  
\begin{equation}
    \vL\sket{\theta, \tilde{\theta}} = \left[-2i\eta(\cos\theta - \cos\Tilde{\theta}) + \gamma_1 e^{i(\theta - \Tilde{\theta})} + \gamma_2 e^{-i(\theta - \Tilde{\theta})} - (\gamma_1 + \gamma_2)\right]\sket{\theta, \Tilde{\theta}},
\end{equation}  
where $\gamma_1$ and $\gamma_2$ are the decay rates associated with $\hat{L}_1$ and $\hat{L}_2$, respectively. Thus, $\vL$ is diagonal in the quasi-momentum basis, meaning that $\sket{\theta,\tilde{\theta}}$ are its eigenvectors. This can alternatively be written as  
\begin{equation}
    \vL\hat{\rho}(\theta,\tilde{\theta}) = \lambda(\theta,\Tilde{\theta})\hat{\rho}(\theta,\Tilde{\theta}),
\end{equation}  
with eigenvalues given by, now separating the real and imaginary parts,  
\begin{equation}
    \lambda(\theta,\Tilde{\theta}) = -(\gamma_1 + \gamma_2)\big(1 - \cos(\theta - \Tilde{\theta})\big) + i\bigg( (\gamma_1 - \gamma_2)\sin(\theta - \Tilde{\theta}) - 2\eta(\cos\theta - \cos\Tilde{\theta})\bigg).
    \label{smeq:olm_eval}
\end{equation}

Equation~(\ref{smeq:olm_eval}) reveals that the stationary states are diagonal in the quasi-momentum basis, meaning $\theta = \Tilde{\theta}$. Furthermore, since $\lambda(\theta,\theta) = 0$, there are no oscillating coherences, implying that a general steady state can be expressed as
\begin{equation} \rho_{ss} = \int_{-\pi/2}^{\pi/2} \rho(\theta,\theta) \ketbra{\theta} d\theta \mbox{.} \end{equation}
Thus, the open evolution effectively suppresses coherences between states in the quasi-momentum basis. Interestingly, this structure of the stationary states holds for all values of $\gamma_1$ and $\gamma_2$.

This behavior contrasts with the corresponding bosonic model, where the Hamiltonian is given by $\hat{H}=\eta(\hat{a}+\hat{a}^\dagger)$ and the jump operators are $\hat{L}_1 = \hat{a}$ and $\hat{L}_2 = \hat{a}^\dagger$. Stability constraints in the bosonic case require $\gamma_1 \geq \gamma_2$, which restricts the steady-state properties. This highlights the crucial role of bosonic algebra in shaping the LFSL model: while an infinite number of steady states always exist in the former model, the latter only permits such behavior in the balanced case $\gamma_1 = \gamma_2$.

We now explore the connection between the presence of an infinite number of steady states and geometrical frustration in the LFSL. Geometrical frustration in closed lattice systems occurs when there exist highly degenerate ground states arising from the underlying lattice geometry, which prevents the system from further minimizing its energy~\cite{wannier1950antiferromagnetism, moessner2006geometrical}. In an open setting, the natural analog to ground states are the steady states of the Liouvillian, that is, the eigenstates corresponding to zero eigenvalues. Therefore, geometrical frustration in an open lattice model implies the existence of an infinite number of steady states.

To determine whether the LFSL of the above model exhibits frustration, we recast the system as a classical lattice model. In many-body settings (bosons), this typically emerges from a mean-field approximation, where onsite operators are replaced by complex numbers, transforming the lattice Hamiltonian into a classical energy functional. Specifically, we assign a complex variable  
\[
\psi_\mathbf{i} = \sqrt{n_\mathbf{i}} e^{i\theta_\mathbf{i}}
\]
to each lattice site $\mathbf{i} = (i,j)$ and optimize the functional $\bar{\mathcal{L}}[n_\mathbf{i},\theta_\mathbf{i}]$. In the traditional framework, $\psi_\mathbf{i}$ represents the local order parameter, where $n_\mathbf{i}$ corresponds to the particle number/density at site $\mathbf{i}$, and $\theta_\mathbf{i}$ denotes its phase. In our case, these quantities correspond to the complex elements of the density operator.  

For a Hamiltonian system, the optimization process amounts to finding the global minimum of the energy. The key idea is that if a minimum can be identified within a unit cell, it also uniquely determines the global minimum. Due to translational invariance, one typically assumes constant densities, $n_\mathbf{i} \equiv n$, and focuses on the "phase locking" induced by tunneling terms and their phases. Frustration occurs when it is impossible to minimize the different terms in the energy functional individually. The anti-ferromagnetic triangular lattice serves as the paradigmatic example, where it is not possible to arrange the spins such that all nearest neighbors are perfectly anti-parallel.

In contrast, for our Liouvillian system, we do not search for a minimum but rather for a solution satisfying $\bar{\mathcal{L}}[n_\mathbf{i},\theta_\mathbf{i}] = 0$, which corresponds to a steady state. As previously discussed, the CPTP property of the Lindblad master equation and the requirement that $\sket{\rho}$ represents a physical state impose constraints on the complex numbers $\psi_\mathbf{i}$. However, to gain insight, we analyze a single unit cell of the LFSL, treat these numbers as independent variables, and assume constant density $n_\mathbf{i} \equiv n$. Hence, we search for a solution of the functional equation with respect to the phase angles $\theta_\mathbf{i}$ only. 

\begin{figure}[!h]
    \centering
    \includegraphics[width=0.25\linewidth]{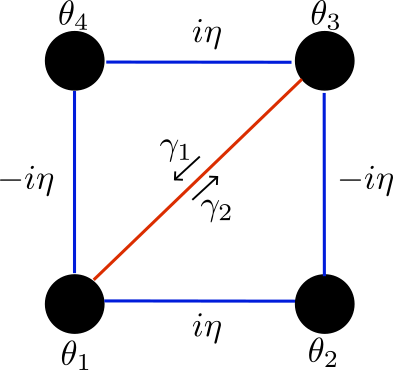}
    \caption{Unit cell of the LFSL corresponding to the open tight-binding model~(\ref{1dlat}). Each site is assigned a phase $\theta_i$, and the relevant tunneling rates are explicitly indicated.}
    \label{fig:olm_unitcell}
\end{figure}

The unit cell with its tunneling rates is depicted in Fig.~\ref{fig:olm_unitcell}. The "unit cell mean-field" Liouvillian functional becomes, up to trivial constants,
\begin{equation}\label{func}
    \bar{\mathcal{L}}[\psi_\mathbf{i}]=i\eta\left(\psi_1^*\psi_2+\psi_2^*\psi_1\right)-i\eta\left(\psi_2^*\psi_3+\psi_3^*\psi_2\right)+i\eta\left(\psi_3^*\psi_4+\psi_4^*\psi_3\right)-i\eta\left(\psi_4^*\psi_1+\psi_1^*\psi_4\right)+\gamma_1\psi_1^*\psi_3+\gamma_2\psi_3^*\psi_1.
\end{equation}
Using the assumption $n_i\equiv n$, and setting $\bar{\mathcal{L}}=0$, we arrive at the following two equations after separating real and imaginary parts:
\begin{equation}\label{angles}
\begin{array}{c}
    \cos(\theta_1-\theta_2)-\cos(\theta_2-\theta_3)+\cos(\theta_3-\theta_4)-\cos(\theta_1-\theta_4)+(\alpha_2-\alpha_1)\sin(\theta_1-\theta_3)=0,\\ \\
    (\alpha_1+\alpha_2)\cos(\theta_1-\theta_3)=0,
    \end{array}
\end{equation}
where $\alpha_{1,2}=\gamma_{1,2}/2\eta$. 

If the high degeneracy of steady states stems from frustration, we expect that we cannot independently identify the different terms of~(\ref{func}) to be zero. Without loss of generality, we can set $\theta_1=0$. From the second equation, we obtain, for example, $\theta_3=\pi/2$. However, one sees that each term of the first equation cannot be made to be zero due to the term $(\alpha_2 - \alpha_1)\sin(\theta_1 - \theta_3) = -(\alpha_2 - \alpha_1)$. In the balanced case $\gamma_1=\gamma_2$ we still have that each term of the first equation cannot be made zero simultaneously. Another way to see that is by checking the case: for the first term to vanish, we require $\theta_1-\theta_2=\pi/2$ (up to additional multiples of $\pi$), and similarly, for the second term, $\theta_2-\theta_3=\pi/2$. This implies that $\theta_1-\theta_3$ must be a multiple of $\pi$, leading to $\cos(\theta_1-\theta_3)\neq0$, and the second equation of~(\ref{angles}) is not fulfilled. Thus, there is no choice of $(\theta_1,\theta_2,\theta_3,\theta_4)$ that simultaneously causes every term in the functional to vanish. We therefore conclude that the steady-state degeneracy is a result of frustration. 

Giving up the constraint that every term of~(\ref{angles}) should simultaneously vanish, we note that for certain parameters $\alpha_{1,2}$, the equations~(\ref{angles}) do not permit real solutions $\theta_i$. One may expect that in such cases it is not legitimate to assume $n_i\equiv n$, and global (classical) solutions, beyond the unit cell, will break the translational invariance of the LFSL.

\end{bibunit}


\begin{thebibliography}{31}
\bibitem [{fta()}]{fta}%
  \BibitemOpen
  \href@noop {} {}\bibinfo {note} {Here, information refers to any parameter
  that characterizes the initial state.}\BibitemShut {Stop}%
\bibitem [{ftb()}]{ftb}%
  \BibitemOpen
  \href@noop {} {}\bibinfo {note} {Throughout this paper, we set $\hbar =
  1$.}\BibitemShut {Stop}%
\end{thebibliography}
\end{document}